\documentclass[10pt,twocolumn,superscriptaddress,floats,showpacs,nobalancelastpage,longbibliography,prb]{revtex4-2}

\usepackage[pdftex]{color}
\usepackage[dvipsnames]{xcolor}

\usepackage{array,longtable}
\newcolumntype{L}{>{\tiny $}p{0.33\columnwidth}<{$}}
\newcolumntype{M}{>{\scriptsize $}p{0.33\columnwidth}<{$}}
\newcolumntype{N}{>{\scriptsize $}p{0.43\columnwidth}<{$}}
\usepackage{amsmath}
\usepackage{amssymb}
\usepackage{amsfonts}
\usepackage{graphicx}
\usepackage{hyperref}
\usepackage{tabularx}
\usepackage{wasysym}

\usepackage{float}
\usepackage{graphics}
\usepackage[caption=false]{subfig}
\usepackage{tikz}
\usepackage{times}
\usepackage{dsfont}
\usepackage{subfig}

\usepackage[normalem]{ulem}

\allowdisplaybreaks[1]

\newif\ifhyper
\hypertrue
\ifhyper
\hypersetup{
   citecolor = {OliveGreen},
   colorlinks = {true}, 
   urlcolor = {RoyalBlue}, 
   linkcolor = {RoyalBlue}
}
\fi

\newenvironment{diagram}
{
\begin{tikzpicture}[baseline = (X.base),every node/.style={scale=0.8},scale=0.45]
}
{
\end{tikzpicture}
}

\begin{document}

\title{Generating function for projected entangled-pair states}

\author{Wei-Lin Tu}
\email{weilintu@keio.jp}
\affiliation{Faculty of Science and Technology, Keio University, 3-14-1 Hiyoshi, Kohoku-ku, Yokohama 223-8522, Japan}

\author{Laurens Vanderstraeten}
\email{laurens.vanderstraeten@ugent.be}
\affiliation{Center for Nonlinear Phenomena and Complex Systems, Université Libre de Bruxelles, Belgium}

\author{Norbert Schuch}
\email{norbert.schuch@univie.ac.at}
\affiliation{University of Vienna, Faculty of Physics, Boltzmanngasse 5, 1090 Wien, Austria}
\affiliation{University of Vienna, Faculty of Mathematics, Oskar-Morgenstern-Platz 1, 1090 Wien, Austria}

\author{Hyun-Yong Lee}
\email{hyunyong@korea.ac.kr}
\affiliation{Division of Display and Semiconductor Physics, Korea University, Sejong 30019, Korea}
\affiliation{Department of Applied Physics, Graduate School, Korea University, Sejong 30019, Korea}
\affiliation{Interdisciplinary Program in E$\cdot$ICT-Culture-Sports Convergence, Korea University, Sejong 30019, Korea}

\author{Naoki Kawashima}
\email{kawashima@issp.u-tokyo.ac.jp}
\affiliation{Institute for Solid State Physics, The University of Tokyo, Kashiwa, Chiba 277-8581, Japan}
\affiliation{Trans-scale Quantum Science Institute, The University of Tokyo, Bunkyo, Tokyo 113-0033, Japan}

\author{Ji-Yao Chen}
\email{chenjiy3@mail.sysu.edu.cn}
\affiliation{Center for Neutron Science and Technology, Guangdong Provincial Key Laboratory of Magnetoelectric Physics and Devices, School of Physics, Sun Yat-sen University, Guangzhou 510275, China}

\date{\today}

\begin{abstract}

Diagrammatic summation is a common bottleneck in modern applications of projected entangled-pair states, especially in computing low-energy excitations of a two-dimensional quantum many-body system. To solve this problem, here we extend the generating function approach for tensor network diagrammatic summation, a scheme previously proposed in the context of matrix product states. Taking the form of a one-particle excitation, we show that the excited state can be computed efficiently in the generating function formalism, which can further be used in evaluating the dynamical structure factor of the system. Our benchmark results for the spin-$1/2$ transverse-field Ising model and Heisenberg model on the square lattice provide a desirable accuracy, showing good agreement with known results. We then study the spin-$1/2$ $J_1$-$J_2$ model on the same lattice and investigate the dynamical properties of the putative gapless spin liquid phase. We conclude with a discussion on generalizations to multi-particle excitations.

\end{abstract}

\maketitle

\section{Introduction}
\label{sec:Introduction}

Strongly correlated quantum systems occupy a central position in condensed matter physics, often triggering exotic behavior at low temperatures. For low-dimensional systems, where quantum effects are pronounced, the entanglement-based tensor network method is now widely recognized as an ideal tool for both analytical and numerical studies of these systems~\cite{Cirac2021, Banuls2023}. Beyond its immense success in exploring ground state properties due to its structure that adheres to the area law, tensor network methods also enable the study of low-energy properties above the ground state, including dynamical correlations and entanglement dynamics. The former are often directly measurable in spectroscopic experiments within condensed matter physics, such as inelastic neutron scattering experiments in quantum magnets~\cite{Banerjee2017}, while out-of-equilibrium properties are accessible in quantum simulators~\cite{Gross2017}. However, in systems extending beyond one spatial dimension, efforts to fully understand low-energy excitations in quantum many-body systems through tensor networks are still in their early stages. Consequently, there is a high demand for an efficient and accurate method to compute low-lying excitations and related dynamical correlation functions in two-dimensional (2D) systems.

Over the past twenty years, projected entangled-pair states (PEPS) have become one of the cornerstones in the study of 2D quantum many-body systems~\cite{Verstraete2004}. Along with a deeper understanding of its mathematical structure and improved numerical recipes, PEPS has been demonstrated to capture the ground states of many classes of two-dimensional phases of matter. These include non-chiral topological states ~\cite{Schuch2010, Williamson2016, Francuz2020, Iqbal2021}, ordered quantum magnets \cite{Corboz2011, Corboz2013a, Corboz2013b, Corboz2014b, Hasik2021, Tu2022, Tu2023}, (chiral) quantum spin liquids~\cite{Schuch2012, Lee2019, Chen2018, Chen2020, Hasik2022b, Xu2023}, and various phases in fermionic systems~\cite{Corboz2014, Zheng2017, Ponsioen2019, Mortier2022, Ponsioen2023a}. Beyond ground state properties, the PEPS toolbox has been expanded to study excited states~\cite{Vanderstraeten2015, Vanderstraeten2019a, Ponsioen2020, Ponsioen2022, Chi2022}, time evolution \cite{Kshetrimayum2017, Czarnik2019b, Dziarmaga2021, McKeever2021, Dziarmaga2022, Lin2022} and finite temperature properties \cite{Czarnik2012, Czarnik2019, Kshetrimayum2019, Poilblanc2021, Jimenez2021, Sinha2022, Vanhecke2023}. Although PEPS-based ground state exploration has reached a level of maturity, research into excited states continues to be a vibrant area of development. Therefore, it is both natural and promising to refine theoretical tools further, leveraging the success with ground states to explore the excitations in 2D systems using PEPS.

One physically motivated way of studying excitations is to use the tensor network generalization~\cite{Ostlund1995, Haegeman2012} of the Feynman-Bijl ansatz \cite{Feynman1953a, Feynman1953b, Feynman1954, Feynman1956} or the single mode approximation \cite{Girvin1986, Arovas1988}, which is a variational ansatz for an excited state in the tangent space of the ground state tensor manifold~\cite{Haegeman2013b, Vanderstraeten2019b}. In this construction, the ground state is perturbed locally by introducing an ``impurity'' tensor to represent a local quasiparticle. By making a momentum superposition of such a local perturbation, one obtains a natural representation of a low-energy excited state. This approach has been successfully applied to build excitations on top of ground states of matrix product state (MPS) form in 1D \cite{Buyens2014, Bera2017, Vanderstraeten2018, ZaunerStauber2018, VanDamme2021, Milsted2022} and quasi-1D \cite{VanDamme2021, Kadow2022, Drescher2022} systems, where contractions can be performed exactly and efficiently. For a 2D quantum many-body system in the thermodynamic limit, this construction leads to the sum of infinitely many copies of PEPS with infinite size, so that computing expectation values or optimizing the variational energy is not straightforward. This can still be achieved through summing a geometric series of channel operators built from the boundary MPS approach~\cite{Vanderstraeten2015, Vanderstraeten2019a} or alternatively through summing diagrams within the corner transfer matrix renormalization group method~\cite{Ponsioen2020, Ponsioen2022}. Nevertheless, the number of tensor diagrams one needs to take care of in these approaches is substantial. Additionally, it is not clear how to apply these approaches to finite-size systems, since both approaches rely on a fixed-point iteration which is designed for an infinite system.

Instead of directly conducting the tensor diagram summation, in Ref.~\onlinecite{Tu2021} some of the authors and collaborators have introduced a set of generating functions for tensor network diagrammatic summations, and applied this scheme to study the excitation spectrum in 1D systems with MPS. The key idea of the generating function is that the relevant tensor diagram summations can be expressed as low-order derivatives of a single diagram. The origin of this scheme can be traced to the fact that interactions in quantum many-body systems are local and the low-energy excited states only contain one or few quasi-particle excitations. Thus, inspired by the generating functional method in quantum field theory, one can introduce a source term in the tensor network and express the excited state as a first-order derivative of a new tensor diagram. With straightforward extensions, the effective Hamiltonian and norm matrices in the variational parameter space of excitations can be obtained, making it simple to evaluate expectation values or optimize the variational parameters.

In this work, we will extend this idea to PEPS in the thermodynamic limit. We begin with a short explanation of the infinite PEPS (iPEPS) ground state ansatz and how to construct low-energy excited states on top of a PEPS ground state in Sec.~\ref{sec:preliminary}. Then we introduce the generating function for PEPS and illustrate how to solve several practical issues in Sec.~\ref{sec:GenFunc}. With impurity tensors obtained through solving an eigenvalue problem, it is straightforward to study the low-energy properties of 2D systems, e.g., the dynamical spin structure factor. In Sec.~\ref{sec:application} we will first benchmark our approach using the well-studied transverse-field Ising model and Heisenberg model on the square lattice, and then apply the method to the more challenging spin-$1/2$ $J_1-J_2$ model on the same lattice. We conclude with some perspectives on generalizing this approach to more general excitations and other settings in Sec.~\ref{sec:conclusion}.

\section{Preliminaries}
\label{sec:preliminary}

\subsection{PEPS as a variational ground state}
\label{subsec:ground_state}

For a 2D quantum many-body system, despite an exponentially large Hilbert space, the ground state of a gapped system obeys the so-called entanglement area law~\cite{Eisert2010, Anshu2022}. This guarantees the validity of the direct extension of MPS for 1D quantum systems to PEPS as a suitable numerical tool for studying ground states of gapped 2D systems. For 2D critical systems, which could violate the entanglement area law, PEPS have also been shown to be a powerful tool which can capture the critical properties via a so-called finite entanglement scaling approach~\cite{Rader2018,  Corboz2018, Vanhecke2022}.

On the square lattice, a PEPS is defined by a set of rank-5 tensors, with one tensor for each site. Using translation symmetry with a suitable unit cell of tensors, PEPS can be defined directly in the thermodynamic limit, giving rise to the so-called iPEPS method. Taking an iPEPS with a one-site unit cell as an example, the wave function is given by the following tensor network form:
\begin{equation}
|\Psi(A)\rangle = 
\begin{diagram}
\draw[line width=0.5mm] (1.5, 2.5) -- (1.5, 2);
\draw[line width=0.5mm] (3, 2.5) -- (3, 2);
\draw[line width=0.5mm] (4.5, 2.5) -- (4.5, 2);
\draw[line width=0.5mm] (6, 2.5) -- (6, 2);
\draw[line width=0.5mm] (7.5, 2.5) -- (7.5, 2);
\draw[line width=0.5mm] (0.5, 1.5) -- (1, 1.5); \draw[color=red!60, very thick](1.5, 1.5) circle (0.5); \draw (1.5, 1.5) node (X) {$A$}; \draw[color=red!60, line width=0.5mm] (1.9, 1.2) -- (2.4, 0.9);
\draw[line width=0.5mm] (2, 1.5) -- (2.5, 1.5); \draw[color=red!60, very thick](3, 1.5) circle (0.5); \draw (3, 1.5) node {$A$}; \draw[color=red!60, line width=0.5mm] (3.4, 1.2) -- (3.9, 0.9);
\draw[line width=0.5mm] (3.5, 1.5) -- (4, 1.5); \draw[color=red!60, very thick](4.5, 1.5) circle (0.5); \draw (4.5, 1.5) node {$A$}; \draw[color=red!60, line width=0.5mm] (4.9, 1.2) -- (5.4, 0.9);
\draw[line width=0.5mm] (5, 1.5) -- (5.5, 1.5); \draw[color=red!60, very thick](6, 1.5) circle (0.5); \draw (6, 1.5) node {$A$}; \draw[color=red!60, line width=0.5mm] (6.4, 1.2) -- (6.9, 0.9);
\draw[line width=0.5mm] (6.5, 1.5) -- (7, 1.5); \draw[color=red!60, very thick](7.5, 1.5) circle (0.5); \draw (7.5, 1.5) node {$A$}; \draw[color=red!60, line width=0.5mm] (7.9, 1.2) -- (8.4, 0.9);
\draw[line width=0.5mm] (8, 1.5) -- (8.5, 1.5);
\draw[line width=0.5mm] (1.5, 1) -- (1.5, 0.5);
\draw[line width=0.5mm] (3, 1) -- (3, 0.5);
\draw[line width=0.5mm] (4.5, 1) -- (4.5, 0.5);
\draw[line width=0.5mm] (6, 1) -- (6, 0.5);
\draw[line width=0.5mm] (7.5, 1) -- (7.5, 0.5);

\draw[line width=0.5mm] (0.5, 0) -- (1, 0); \draw[color=red!60, very thick](1.5, 0) circle (0.5); \draw (1.5, 0) node (X) {$A$}; \draw[color=red!60, line width=0.5mm] (1.9, -0.3) -- (2.4, -0.6);
\draw[line width=0.5mm] (2, 0) -- (2.5, 0); \draw[color=red!60, very thick](3, 0) circle (0.5); \draw (3, 0) node {$A$}; \draw[color=red!60, line width=0.5mm] (3.4, -0.3) -- (3.9, -0.6);
\draw[line width=0.5mm] (3.5, 0) -- (4, 0); \draw[color=red!60, very thick](4.5, 0) circle (0.5); \draw (4.5, 0) node {$A$}; \draw[color=red!60, line width=0.5mm] (4.9, -0.3) -- (5.4, -0.6);
\draw[line width=0.5mm] (5, 0) -- (5.5, 0); \draw[color=red!60, very thick](6, 0) circle (0.5); \draw (6, 0) node {$A$}; \draw[color=red!60, line width=0.5mm] (6.4, -0.3) -- (6.9, -0.6);
\draw[line width=0.5mm] (6.5, 0) -- (7, 0); \draw[color=red!60, very thick](7.5, 0) circle (0.5); \draw (7.5, 0) node {$A$}; \draw[color=red!60, line width=0.5mm] (7.9, -0.3) -- (8.4, -0.6);
\draw[line width=0.5mm] (8, 0) -- (8.5, 0);
\draw[line width=0.5mm] (1.5, -0.5) -- (1.5, -1);
\draw[line width=0.5mm] (3, -0.5) -- (3, -1);
\draw[line width=0.5mm] (4.5, -0.5) -- (4.5, -1);
\draw[line width=0.5mm] (6, -0.5) -- (6, -1);
\draw[line width=0.5mm] (7.5, -0.5) -- (7.5, -1);

\draw[line width=0.5mm] (0.5, -1.5) -- (1, -1.5); \draw[color=red!60, very thick](1.5, -1.5) circle (0.5); \draw (1.5, -1.5) node (X) {$A$}; \draw[color=red!60, line width=0.5mm] (1.9, -1.8) -- (2.4, -2.1);
\draw[line width=0.5mm] (2, -1.5) -- (2.5, -1.5); \draw[color=red!60, very thick](3, -1.5) circle (0.5); \draw (3, -1.5) node {$A$}; \draw[color=red!60, line width=0.5mm] (3.4, -1.8) -- (3.9, -2.1);
\draw[line width=0.5mm] (3.5, -1.5) -- (4, -1.5); \draw[color=red!60, very thick](4.5, -1.5) circle (0.5); \draw (4.5, -1.5) node {$A$}; \draw[color=red!60, line width=0.5mm] (4.9, -1.8) -- (5.4, -2.1);
\draw[line width=0.5mm] (5, -1.5) -- (5.5, -1.5); \draw[color=red!60, very thick](6, -1.5) circle (0.5); \draw (6, -1.5) node {$A$}; \draw[color=red!60, line width=0.5mm] (6.4, -1.8) -- (6.9, -2.1);
\draw[line width=0.5mm] (6.5, -1.5) -- (7, -1.5); \draw[color=red!60, very thick](7.5, -1.5) circle (0.5); \draw (7.5, -1.5) node {$A$}; \draw[color=red!60, line width=0.5mm] (7.9, -1.8) -- (8.4, -2.1);
\draw[line width=0.5mm] (8, -1.5) -- (8.5, -1.5);
\draw[line width=0.5mm] (1.5, -2) -- (1.5, -2.5);
\draw[line width=0.5mm] (3, -2) -- (3, -2.5);
\draw[line width=0.5mm] (4.5, -2) -- (4.5, -2.5);
\draw[line width=0.5mm] (6, -2) -- (6, -2.5);
\draw[line width=0.5mm] (7.5, -2) -- (7.5, -2.5);
\draw (0, -1.5) node {$\ldots$};
\draw (9, -1.5) node {$\ldots$};
\draw (4.5, 3.2) node {$\vdots$};
\draw (4.5, -5.8) node {$\vdots$};

\draw[line width=0.5mm] (0.5, -3) -- (1, -3); \draw[color=red!60, very thick](1.5, -3) circle (0.5); \draw (1.5, -3) node {$A$}; \draw[color=red!60, line width=0.5mm] (1.9, -3.3) -- (2.4, -3.6);
\draw[line width=0.5mm] (2, -3) -- (2.5, -3); \draw[color=red!60, very thick](3, -3) circle (0.5); \draw (3, -3) node {$A$}; \draw[color=red!60, line width=0.5mm] (3.4, -3.3) -- (3.9, -3.6);
\draw[line width=0.5mm] (3.5, -3) -- (4, -3); \draw[color=red!60, very thick](4.5, -3) circle (0.5); \draw (4.5, -3) node {$A$}; \draw[color=red!60, line width=0.5mm] (4.9, -3.3) -- (5.4, -3.6);
\draw[line width=0.5mm] (5, -3) -- (5.5, -3); \draw[color=red!60, very thick](6, -3) circle (0.5); \draw (6, -3) node {$A$}; \draw[color=red!60, line width=0.5mm] (6.4, -3.3) -- (6.9, -3.6);
\draw[line width=0.5mm] (6.5, -3) -- (7, -3); \draw[color=red!60, very thick](7.5, -3) circle (0.5); \draw (7.5, -3) node {$A$}; \draw[color=red!60, line width=0.5mm] (7.9, -3.3) -- (8.4, -3.6);
\draw[line width=0.5mm] (8, -3) -- (8.5, -3);
\draw[line width=0.5mm] (1.5, -3.5) -- (1.5, -4);
\draw[line width=0.5mm] (3, -3.5) -- (3, -4);
\draw[line width=0.5mm] (4.5, -3.5) -- (4.5, -4);
\draw[line width=0.5mm] (6, -3.5) -- (6, -4);
\draw[line width=0.5mm] (7.5, -3.5) -- (7.5, -4);

\draw[line width=0.5mm] (0.5, -4.5) -- (1, -4.5); \draw[color=red!60, very thick](1.5, -4.5) circle (0.5); \draw (1.5, -4.5) node {$A$}; \draw[color=red!60, line width=0.5mm] (1.9, -4.8) -- (2.4, -5.1);
\draw[line width=0.5mm] (2, -4.5) -- (2.5, -4.5); \draw[color=red!60, very thick](3, -4.5) circle (0.5); \draw (3, -4.5) node {$A$}; \draw[color=red!60, line width=0.5mm] (3.4, -4.8) -- (3.9, -5.1);
\draw[line width=0.5mm] (3.5, -4.5) -- (4, -4.5); \draw[color=red!60, very thick](4.5, -4.5) circle (0.5); \draw (4.5, -4.5) node {$A$}; \draw[color=red!60, line width=0.5mm] (4.9, -4.8) -- (5.4, -5.1);
\draw[line width=0.5mm] (5, -4.5) -- (5.5, -4.5); \draw[color=red!60, very thick](6, -4.5) circle (0.5); \draw (6, -4.5) node {$A$}; \draw[color=red!60, line width=0.5mm] (6.4, -4.8) -- (6.9, -5.1);
\draw[line width=0.5mm] (6.5, -4.5) -- (7, -4.5); \draw[color=red!60, very thick](7.5, -4.5) circle (0.5); \draw (7.5, -4.5) node {$A$}; \draw[color=red!60, line width=0.5mm] (7.9, -4.8) -- (8.4, -5.1);
\draw[line width=0.5mm] (8, -4.5) -- (8.5, -4.5);
\draw[line width=0.5mm] (1.5, -5) -- (1.5, -5.5);
\draw[line width=0.5mm] (3, -5) -- (3, -5.5);
\draw[line width=0.5mm] (4.5, -5) -- (4.5, -5.5);
\draw[line width=0.5mm] (6, -5) -- (6, -5.5);
\draw[line width=0.5mm] (7.5, -5) -- (7.5, -5.5);
\end{diagram},
\label{eq:uniformPEPS}
\end{equation}
where the black (red) bonds represent the virtual (physical) indices with bond dimension $D$ ($d$). The accuracy of approximating the ground state of a 2D quantum system with the PEPS ansatz can be controlled by the virtual bond dimension $D$.

Although in general exact contraction of PEPS in the thermodynamic limit is not possible, various controlled approximate contraction schemes for PEPS have been developed, including the boundary MPS approach~\cite{Jordan2008, Fishman2018, Vanderstraeten2022}, the tensor renormalization group and its various generalizations~\cite{Levin2007, Xie2009, Xie2012}, and the corner transfer matrix renormalization group (CTMRG) method~\cite{Baxter1978, Nishino1996, Orus2009, Corboz2014}. In this work we will use the CTMRG method, where the basic idea is to approximate the surroundings of every site using a set of tensors with finite bond dimension, typically denoted as the environment bond dimension $\chi$. As an illustration, we first trace out the physical index of $A$ and $\bar{A}$, where $\bar{A}$ represents the complex conjugate of tensor $A$, creating the norm of the PEPS wave function:
\begin{equation}
\langle\Psi(A)|\Psi(A)\rangle = 
\begin{diagram}
\draw[line width=1mm] (1.5, 1) -- (1.5, 0.5);
\draw[line width=1mm] (3, 1) -- (3, 0.5);
\draw[line width=1mm] (4.5, 1) -- (4.5, 0.5);
\draw[line width=1mm] (0.5, 0) -- (1, 0); \fill[color=red!60, very thick](1.5, 0) circle (0.5);
\draw[line width=1mm] (2, 0) -- (2.5, 0); \fill[color=red!60, very thick](3, 0) circle (0.5);
\draw[line width=1mm] (3.5, 0) -- (4, 0); \fill[color=red!60, very thick](4.5, 0) circle (0.5);
\draw[line width=1mm] (5, 0) -- (5.5, 0);
\draw[line width=1mm] (1.5, -0.5) -- (1.5, -1);
\draw[line width=1mm] (3, -0.5) -- (3, -1);
\draw[line width=1mm] (4.5, -0.5) -- (4.5, -1);

\draw[line width=1mm] (0.5, -1.5) -- (1, -1.5); \fill[color=red!60, very thick](1.5, -1.5) circle (0.5); \draw (1.5, -1.5) node (X) {};
\draw[line width=1mm] (2, -1.5) -- (2.5, -1.5); \fill[color=red!60, very thick](3, -1.5) circle (0.5);
\draw[line width=1mm] (3.5, -1.5) -- (4, -1.5); \fill[color=red!60, very thick](4.5, -1.5) circle (0.5);
\draw[line width=1mm] (5, -1.5) -- (5.5, -1.5);
\draw[line width=1mm] (1.5, -2) -- (1.5, -2.5);
\draw[line width=1mm] (3, -2) -- (3, -2.5);
\draw[line width=1mm] (4.5, -2) -- (4.5, -2.5);

\draw[line width=1mm] (0.5, -3) -- (1, -3); \fill[color=red!60, very thick](1.5, -3) circle (0.5);
\draw[line width=1mm] (2, -3) -- (2.5, -3); \fill[color=red!60, very thick](3, -3) circle (0.5);
\draw[line width=1mm] (3.5, -3) -- (4, -3); \fill[color=red!60, very thick](4.5, -3) circle (0.5);
\draw[line width=1mm] (5, -3) -- (5.5, -3);
\draw[line width=1mm] (1.5, -3.5) -- (1.5, -4);
\draw[line width=1mm] (3, -3.5) -- (3, -4);
\draw[line width=1mm] (4.5, -3.5) -- (4.5, -4);
\draw (0, -1.5) node {$\ldots$};
\draw (6, -1.5) node {$\ldots$};
\draw (3, 1.7) node {$\vdots$};
\draw (3, -4.3) node {$\vdots$};
\end{diagram},
\label{eq:doublelayer}
\end{equation}
where
\begin{equation}
\begin{diagram}
\draw[line width=1mm] (1.5, -0.3) -- (1.5, -0.8);
\draw[line width=1mm] (0.5, -1.3) -- (1, -1.3); \fill[color=red!60, very thick](1.5, -1.3) circle (0.5);
\draw[line width=1mm] (2, -1.3) -- (2.5, -1.3);
\draw[line width=1mm] (1.5, -1.8) -- (1.5, -2.3);
\end{diagram}
=
\begin{diagram}
\draw[line width=0.5mm] (1.1, 1.0) -- (1.5, 0.4);
\draw[line width=0.5mm] (0.5, 0) -- (1.3, 0); \draw[color=red!60, very thick](1.8, 0) circle (0.5); \draw (1.8, 0) node {$A$};
\draw[line width=0.5mm] (2.3, 0) -- (3.1, 0);
\draw[line width=0.5mm] (2.1, -0.4) -- (2.5, -1);
\draw[color=red!60, line width=0.5mm] (1.8, -0.5) -- (1.8, -2);
\draw[line width=0.5mm] (1.1, -1.5) -- (1.5, -2.1);
\draw[line width=0.5mm] (0.5, -2.5) -- (1.3, -2.5); \draw[color=red!60, very thick](1.8, -2.5) circle (0.5); \draw (1.8, -2.5) node {$\bar{A}$};
\draw[line width=0.5mm] (2.3, -2.5) -- (3.1, -2.5);
\draw[line width=0.5mm] (2.1, -2.9) -- (2.5, -3.5);
\end{diagram}.
\label{eq:phystrace}
\end{equation}
The thick black bonds in Eq.~\eqref{eq:phystrace} thus have dimension $D^2$. Through the CTMRG procedure, one can obtain a series of environment tensors that serve as the approximation of all surrounding tensors:
\begin{equation}
\begin{diagram}
\draw[line width=1mm] (1.5, 1) -- (1.5, 0.5);
\draw[line width=1mm] (3, 1) -- (3, 0.5);
\draw[line width=1mm] (4.5, 1) -- (4.5, 0.5);
\draw[line width=1mm] (0.5, 0) -- (1, 0); \fill[color=red!60, very thick](1.5, 0) circle (0.5);
\draw[line width=1mm] (2, 0) -- (2.5, 0); \fill[color=red!60, very thick](3, 0) circle (0.5);
\draw[line width=1mm] (3.5, 0) -- (4, 0); \fill[color=red!60, very thick](4.5, 0) circle (0.5);
\draw[line width=1mm] (5, 0) -- (5.5, 0);
\draw[line width=1mm] (1.5, -0.5) -- (1.5, -1);
\draw[line width=1mm] (3, -0.5) -- (3, -1);
\draw[line width=1mm] (4.5, -0.5) -- (4.5, -1);

\draw[line width=1mm] (0.5, -1.5) -- (1, -1.5); \fill[color=red!60, very thick](1.5, -1.5) circle (0.5); \draw (1.5, -1.5) node (X) {};
\draw[line width=1mm] (2, -1.5) -- (2.5, -1.5); \fill[color=red!60, very thick](3, -1.5) circle (0.5);
\draw[line width=1mm] (3.5, -1.5) -- (4, -1.5); \fill[color=red!60, very thick](4.5, -1.5) circle (0.5);
\draw[line width=1mm] (5, -1.5) -- (5.5, -1.5);
\draw[line width=1mm] (1.5, -2) -- (1.5, -2.5);
\draw[line width=1mm] (3, -2) -- (3, -2.5);
\draw[line width=1mm] (4.5, -2) -- (4.5, -2.5);

\draw[line width=1mm] (0.5, -3) -- (1, -3); \fill[color=red!60, very thick](1.5, -3) circle (0.5);
\draw[line width=1mm] (2, -3) -- (2.5, -3); \fill[color=red!60, very thick](3, -3) circle (0.5);
\draw[line width=1mm] (3.5, -3) -- (4, -3); \fill[color=red!60, very thick](4.5, -3) circle (0.5);
\draw[line width=1mm] (5, -3) -- (5.5, -3);
\draw[line width=1mm] (1.5, -3.5) -- (1.5, -4);
\draw[line width=1mm] (3, -3.5) -- (3, -4);
\draw[line width=1mm] (4.5, -3.5) -- (4.5, -4);
\draw (0, -1.5) node {$\ldots$};
\draw (6, -1.5) node {$\ldots$};
\draw (3, 1.7) node {$\vdots$};
\draw (3, -4.3) node {$\vdots$};
\end{diagram}
\approx
\begin{diagram}
\draw[color=black!60, very thick](2, 0.5) rectangle (1, -0.5); \draw (1.5, 0) node {$C_1$};
\draw[color=black!30!green, line width=1mm] (2, 0) -- (2.5, 0); \draw[color=black!60, very thick](3.5, 0.7) rectangle (2.5, -0.5); \draw (3, 0) node {$T_1$};
\draw[color=black!30!green, line width=1mm] (4, 0) -- (3.5, 0); \draw[color=black!60, very thick](5, 0.5) rectangle (4, -0.5); \draw (4.5, 0) node {$C_2$};
\draw[color=black!30!green, line width=1mm] (1.5, -0.5) -- (1.5, -1);
\draw[line width=1mm] (3, -0.5) -- (3, -1);
\draw[color=black!30!green, line width=1mm] (4.5, -0.5) -- (4.5, -1);

\draw[color=black!60, very thick](2, -1) rectangle (0.8, -2); \draw (1.5, -1.5) node (X) {$T_4$};
\draw[line width=1mm] (2, -1.5) -- (2.5, -1.5); \fill[color=red!60, very thick](3, -1.5) circle (0.5);
\draw[line width=1mm] (3.5, -1.5) -- (4, -1.5); \draw[color=black!60, very thick](4, -1) rectangle (5.2, -2); \draw (4.5, -1.5) node (X) {$T_2$};
\draw[color=black!30!green, line width=1mm] (1.5, -2) -- (1.5, -2.5);
\draw[line width=1mm] (3, -2) -- (3, -2.5);
\draw[color=black!30!green, line width=1mm] (4.5, -2) -- (4.5, -2.5);

\draw[color=black!60, very thick](2, -2.5) rectangle (1, -3.5); \draw (1.5, -3) node {$C_4$};
\draw[color=black!30!green, line width=1mm] (2, -3) -- (2.5, -3); \draw[color=black!60, very thick](3.5, -2.5) rectangle (2.5, -3.7); \draw (3, -3) node {$T_3$};
\draw[color=black!30!green, line width=1mm] (4, -3) -- (3.5, -3); \draw[color=black!60, very thick](5, -2.5) rectangle (4, -3.5); \draw (4.5, -3) node {$C_3$};
\end{diagram}\;.
\label{eq:CTMRG}
\end{equation}
Here, the $C_i$ tensors are the corner tensors and the $T_i$'s represent the edge tensors ($i=1,2,3,4$). The green bonds~(bonds that connect $C$ and $T$ tensors) have dimension $\chi$ and typically one would choose $\chi\geq D^2$ to ensure the accuracy. Note that, to obtain the environment tensors, one would need to introduce projectors to truncate the growing environment bond dimension in each CTMRG iteration. Apart from the ground state calculation, we will show that the environment tensors and the projectors will also play important roles in the calculation of generating functions for excited states.

Using environment tensors, the reduced density matrix of a local region can be constructed and observables can be evaluated. E.g., given a local Hamiltonian of the form $H=\sum_i h_i$, the energy density can be calculated as:
\begin{equation}
\label{eq:energy}
    e=\frac{\langle\Psi(A)|h_i|\Psi(A)\rangle}{\langle\Psi(A)|\Psi(A)\rangle} \;.
\end{equation}
Note that, when computing Eq.~\eqref{eq:energy}, the approximations due to projectors in CTMRG are the same for the numerator and denominator. This point will be illuminating when considering the contraction scheme for generating functions of PEPS.

The remaining task is to optimize the ground state tensor $A$ to minimize the variational energy. Conventionally, this can be achieved through imaginary time evolution, where the local Hamiltonian is used to construct Suzuki-Trotter gates, and after a sufficiently long imaginary time evolution a well-approximated ground state ansatz can be obtained~\cite{Jiang2008, Jordan2008, Phien2015}. However, the initial ansatz for imaginary time evolution tends to significantly affect the final outcome, hindering the simulation without any tentative knowledge of the target model and sometimes causing a serious issue of getting trapped in the local minima of Hilbert space. Additionally, due to the approximations involved in truncating the PEPS after applying Suzuki-Trotter gates, the imaginary time evolution does not always converge to the variationally optimal PEPS tensor.

Instead of doing imaginary time evolution, minimizing energy through the energy gradient, i.e., variational optimization, has been proven to be a more accurate optimization scheme \cite{Vanderstraeten2016, Corboz2016}. When the number of variational parameters is small (for example, when the PEPS tensors are strongly constrained by symmetries), variational optimization can be done using a simple finite difference approach to compute the energy gradient~\cite{Poilblanc2017, Poilblanc2017b, Chen2018, Chen2020}. For generic tensors, a direct evaluation of the gradient requires a summation of a large number of tensor diagrams~\cite{Corboz2016, Vanderstraeten2016, Vanderstraeten2022}, but the numerical derivatives can be obtained through automatic differentiation (AD)~\cite{Liao2019}. By storing the computational graph of PEPS~\cite{Hasik2021, Tu2022} in the memory during the forward-mode calculation, the numerically exact gradients can be evaluated through the back propagation~\cite{Rigo2022}. We then make use of those gradients to further optimize our tensors in order to lower the variational energy.

In the following simulations, we have checked that for each bond dimension $D$, the chosen value of $\chi$ is either large enough and the results barely change when further increasing $\chi$, or the largest $\chi$ within computational reach. In search of the ground state, we have imposed $C_{4v}$ symmetry of the square lattice on the local tensor, since the models we studied do not break this symmetry. In the actual numerical computation, the ground state optimization is carried out using the~\textsl{peps-torch} package~\cite{peps-torch}, where the energy gradient is obtained using the back propagation of AD provided in PyTorch. For the excited state computation (see later sections), we have written our codes in Python and used PyTorch for back propagation. To lower the bar for PEPS practitioners, we have made a sample code publicly available~\footnote{See \href{https://github.com/weilintu/peps-excitation}{https://github.com/weilintu/peps-excitation} for code implementation in PyTorch}.

\subsection{Quasiparticle excitation with PEPS}
\label{subsec:excited_state}

The excited state for a given many-body Hamiltonian can be thought of as a quasiparticle excitation on top of the ground state~\cite{Haegeman2013a}. As mentioned in the introduction, in real space such an excitation can be approximated through the tensor network version of the single mode approximation, where we replace one ground state tensor with some ``impurity'' tensor at a certain location~\footnote{Considering larger patches is also possible, although the computational cost would also be higher.}. We then sum up all different tensor network configurations generated by the translation operator, taking into account the corresponding phase factors to construct an eigenstate of the translation operator with a given momentum.

With the ground state expressed as a one-site translation invariant iPEPS, the one-particle excitation with PEPS can be written as:
\begin{equation}
|\Phi_\textbf{k}(B)\rangle=\sum_{\textbf{r}=(m,n)}\mathrm{e}^{-i\textbf{k}\cdot\textbf{r}}\hat{T}_x^m \hat{T}_y^n
\begin{diagram}
\draw[line width=0.5mm] (1.5, 2.5) -- (1.5, 2);
\draw[line width=0.5mm] (3, 2.5) -- (3, 2);
\draw[line width=0.5mm] (4.5, 2.5) -- (4.5, 2);
\draw[line width=0.5mm] (6, 2.5) -- (6, 2);
\draw[line width=0.5mm] (7.5, 2.5) -- (7.5, 2);
\draw[line width=0.5mm] (0.5, 1.5) -- (1, 1.5); \draw[color=red!60, very thick](1.5, 1.5) circle (0.5); \draw (1.5, 1.5) node (X) {$A$}; \draw[color=red!60, line width=0.5mm] (1.9, 1.2) -- (2.4, 0.9);
\draw[line width=0.5mm] (2, 1.5) -- (2.5, 1.5); \draw[color=red!60, very thick](3, 1.5) circle (0.5); \draw (3, 1.5) node {$A$}; \draw[color=red!60, line width=0.5mm] (3.4, 1.2) -- (3.9, 0.9);
\draw[line width=0.5mm] (3.5, 1.5) -- (4, 1.5); \draw[color=red!60, very thick](4.5, 1.5) circle (0.5); \draw (4.5, 1.5) node {$A$}; \draw[color=red!60, line width=0.5mm] (4.9, 1.2) -- (5.4, 0.9);
\draw[line width=0.5mm] (5, 1.5) -- (5.5, 1.5); \draw[color=red!60, very thick](6, 1.5) circle (0.5); \draw (6, 1.5) node {$A$}; \draw[color=red!60, line width=0.5mm] (6.4, 1.2) -- (6.9, 0.9);
\draw[line width=0.5mm] (6.5, 1.5) -- (7, 1.5); \draw[color=red!60, very thick](7.5, 1.5) circle (0.5); \draw (7.5, 1.5) node {$A$}; \draw[color=red!60, line width=0.5mm] (7.9, 1.2) -- (8.4, 0.9);
\draw[line width=0.5mm] (8, 1.5) -- (8.5, 1.5);
\draw[line width=0.5mm] (1.5, 1) -- (1.5, 0.5);
\draw[line width=0.5mm] (3, 1) -- (3, 0.5);
\draw[line width=0.5mm] (4.5, 1) -- (4.5, 0.5);
\draw[line width=0.5mm] (6, 1) -- (6, 0.5);
\draw[line width=0.5mm] (7.5, 1) -- (7.5, 0.5);

\draw[line width=0.5mm] (0.5, 0) -- (1, 0); \draw[color=red!60, very thick](1.5, 0) circle (0.5); \draw (1.5, 0) node (X) {$A$}; \draw[color=red!60, line width=0.5mm] (1.9, -0.3) -- (2.4, -0.6);
\draw[line width=0.5mm] (2, 0) -- (2.5, 0); \draw[color=red!60, very thick](3, 0) circle (0.5); \draw (3, 0) node {$A$}; \draw[color=red!60, line width=0.5mm] (3.4, -0.3) -- (3.9, -0.6);
\draw[line width=0.5mm] (3.5, 0) -- (4, 0); \draw[color=red!60, very thick](4.5, 0) circle (0.5); \draw (4.5, 0) node {$A$}; \draw[color=red!60, line width=0.5mm] (4.9, -0.3) -- (5.4, -0.6);
\draw[line width=0.5mm] (5, 0) -- (5.5, 0); \draw[color=red!60, very thick](6, 0) circle (0.5); \draw (6, 0) node {$A$}; \draw[color=red!60, line width=0.5mm] (6.4, -0.3) -- (6.9, -0.6);
\draw[line width=0.5mm] (6.5, 0) -- (7, 0); \draw[color=red!60, very thick](7.5, 0) circle (0.5); \draw (7.5, 0) node {$A$}; \draw[color=red!60, line width=0.5mm] (7.9, -0.3) -- (8.4, -0.6);
\draw[line width=0.5mm] (8, 0) -- (8.5, 0);
\draw[line width=0.5mm] (1.5, -0.5) -- (1.5, -1);
\draw[line width=0.5mm] (3, -0.5) -- (3, -1);
\draw[line width=0.5mm] (4.5, -0.5) -- (4.5, -1);
\draw[line width=0.5mm] (6, -0.5) -- (6, -1);
\draw[line width=0.5mm] (7.5, -0.5) -- (7.5, -1);

\draw[line width=0.5mm] (0.5, -1.5) -- (1, -1.5); \draw[color=red!60, very thick](1.5, -1.5) circle (0.5); \draw (1.5, -1.5) node (X) {$A$}; \draw[color=red!60, line width=0.5mm] (1.9, -1.8) -- (2.4, -2.1);
\draw[line width=0.5mm] (2, -1.5) -- (2.5, -1.5); \draw[color=red!60, very thick](3, -1.5) circle (0.5); \draw (3, -1.5) node {$A$}; \draw[color=red!60, line width=0.5mm] (3.4, -1.8) -- (3.9, -2.1);
\draw[line width=0.5mm] (3.5, -1.5) -- (4, -1.5); \draw[color=blue!60, very thick](4.5, -1.5) circle (0.5); \draw (4.5, -1.5) node {$B$}; \draw[color=blue!60, line width=0.5mm] (4.9, -1.8) -- (5.4, -2.1);
\draw[line width=0.5mm] (5, -1.5) -- (5.5, -1.5); \draw[color=red!60, very thick](6, -1.5) circle (0.5); \draw (6, -1.5) node {$A$}; \draw[color=red!60, line width=0.5mm] (6.4, -1.8) -- (6.9, -2.1);
\draw[line width=0.5mm] (6.5, -1.5) -- (7, -1.5); \draw[color=red!60, very thick](7.5, -1.5) circle (0.5); \draw (7.5, -1.5) node {$A$}; \draw[color=red!60, line width=0.5mm] (7.9, -1.8) -- (8.4, -2.1);
\draw[line width=0.5mm] (8, -1.5) -- (8.5, -1.5);
\draw[line width=0.5mm] (1.5, -2) -- (1.5, -2.5);
\draw[line width=0.5mm] (3, -2) -- (3, -2.5);
\draw[line width=0.5mm] (4.5, -2) -- (4.5, -2.5);
\draw[line width=0.5mm] (6, -2) -- (6, -2.5);
\draw[line width=0.5mm] (7.5, -2) -- (7.5, -2.5);

\draw[line width=0.5mm] (0.5, -3) -- (1, -3); \draw[color=red!60, very thick](1.5, -3) circle (0.5); \draw (1.5, -3) node {$A$}; \draw[color=red!60, line width=0.5mm] (1.9, -3.3) -- (2.4, -3.6);
\draw[line width=0.5mm] (2, -3) -- (2.5, -3); \draw[color=red!60, very thick](3, -3) circle (0.5); \draw (3, -3) node {$A$}; \draw[color=red!60, line width=0.5mm] (3.4, -3.3) -- (3.9, -3.6);
\draw[line width=0.5mm] (3.5, -3) -- (4, -3); \draw[color=red!60, very thick](4.5, -3) circle (0.5); \draw (4.5, -3) node {$A$}; \draw[color=red!60, line width=0.5mm] (4.9, -3.3) -- (5.4, -3.6);
\draw[line width=0.5mm] (5, -3) -- (5.5, -3); \draw[color=red!60, very thick](6, -3) circle (0.5); \draw (6, -3) node {$A$}; \draw[color=red!60, line width=0.5mm] (6.4, -3.3) -- (6.9, -3.6);
\draw[line width=0.5mm] (6.5, -3) -- (7, -3); \draw[color=red!60, very thick](7.5, -3) circle (0.5); \draw (7.5, -3) node {$A$}; \draw[color=red!60, line width=0.5mm] (7.9, -3.3) -- (8.4, -3.6);
\draw[line width=0.5mm] (8, -3) -- (8.5, -3);
\draw[line width=0.5mm] (1.5, -3.5) -- (1.5, -4);
\draw[line width=0.5mm] (3, -3.5) -- (3, -4);
\draw[line width=0.5mm] (4.5, -3.5) -- (4.5, -4);
\draw[line width=0.5mm] (6, -3.5) -- (6, -4);
\draw[line width=0.5mm] (7.5, -3.5) -- (7.5, -4);

\draw[line width=0.5mm] (0.5, -4.5) -- (1, -4.5); \draw[color=red!60, very thick](1.5, -4.5) circle (0.5); \draw (1.5, -4.5) node {$A$}; \draw[color=red!60, line width=0.5mm] (1.9, -4.8) -- (2.4, -5.1);
\draw[line width=0.5mm] (2, -4.5) -- (2.5, -4.5); \draw[color=red!60, very thick](3, -4.5) circle (0.5); \draw (3, -4.5) node {$A$}; \draw[color=red!60, line width=0.5mm] (3.4, -4.8) -- (3.9, -5.1);
\draw[line width=0.5mm] (3.5, -4.5) -- (4, -4.5); \draw[color=red!60, very thick](4.5, -4.5) circle (0.5); \draw (4.5, -4.5) node {$A$}; \draw[color=red!60, line width=0.5mm] (4.9, -4.8) -- (5.4, -5.1);
\draw[line width=0.5mm] (5, -4.5) -- (5.5, -4.5); \draw[color=red!60, very thick](6, -4.5) circle (0.5); \draw (6, -4.5) node {$A$}; \draw[color=red!60, line width=0.5mm] (6.4, -4.8) -- (6.9, -5.1);
\draw[line width=0.5mm] (6.5, -4.5) -- (7, -4.5); \draw[color=red!60, very thick](7.5, -4.5) circle (0.5); \draw (7.5, -4.5) node {$A$}; \draw[color=red!60, line width=0.5mm] (7.9, -4.8) -- (8.4, -5.1);
\draw[line width=0.5mm] (8, -4.5) -- (8.5, -4.5);
\draw[line width=0.5mm] (1.5, -5) -- (1.5, -5.5);
\draw[line width=0.5mm] (3, -5) -- (3, -5.5);
\draw[line width=0.5mm] (4.5, -5) -- (4.5, -5.5);
\draw[line width=0.5mm] (6, -5) -- (6, -5.5);
\draw[line width=0.5mm] (7.5, -5) -- (7.5, -5.5);
\end{diagram},
\label{eq:excitation}
\end{equation}
where $\hat{T}_x$ $(\hat{T}_y)$ represents the translation operator in the $x$ $(y)$-direction with its eigenvalue being $\mathrm{e}^{ik_x}$ $(\mathrm{e}^{ik_y})$, and $B$ is the impurity tensor to be determined.

Due to the linear dependence of excited states on the impurity tensor, the variational optimization boils down to a generalized eigenvalue problem:
\begin{equation}  \label{eq:eigenvalue}
\textbf{H}_{\mu\nu}B_{\nu}=E \textbf{N}_{\mu\nu} B_{\nu},
\end{equation}
where $\textbf{H}$ and $\textbf{N}$ are the effective Hamiltonian and norm matrix in the variational space. With translation invariance, $\textbf{N}$ can be obtained by hollowing out the $\bar{A}$ tensor in the center of the bra layer, and the summation goes over every tensor graph with an empty site in the ket layer. Using channel operators with boundary MPS \cite{Vanderstraeten2015}, this sum can be carried out at the same computational cost as ground state computation. A slightly different summation scheme using CTMRG has also been proposed in Ref.~\cite{Ponsioen2020}. Similarly, the effective Hamiltonian $\textbf{H}$ can also be computed, where an additional summation for the Hamiltonian operator appears. These summations are the main bottleneck of studying excitations using the quasi-particle ansatz.

Because of the gauge freedom in the PEPS representation~(Eq.~\eqref{eq:uniformPEPS}) in terms of the local tensor $A$, the linear subspace of excitations contains a few zero modes \cite{Vanderstraeten2015}. These are reflected as zero eigenvalues in the norm matrix and should be projected out to make the above eigenvalue equation well-conditioned. We will come back to the conditioning of $\textbf{N}$ below.

\section{The generating function}
\label{sec:GenFunc}

\subsection{Algorithm}
\label{subsec:algorithm}

The issue of summing up many diagrams also appears in the MPS study of the one-dimensional quantum system, where some of the authors and collaborators have proposed generating functions to tackle this problem~\cite{Tu2021}. Now we adopt this method in the current setting of  PEPS in the thermodynamic limit. The basic idea of generating function is that the sum of extensive tensor diagrams can be expressed as a low-order derivative of a new tensor diagram. This is possible for excited states due to the fact that tensor diagrams only differ by the location of the impurity tensor.

For iPEPS, following Ref.~\onlinecite{Tu2021}, we define the generating function for the one-particle excited state as:
\begin{equation}
|G_\Phi(\lambda)\rangle=
\begin{diagram}
\draw[line width=0.5mm] (1.5, 2.5) -- (1.5, 2);
\draw[line width=0.5mm] (3, 2.5) -- (3, 2);
\draw[line width=0.5mm] (4.5, 2.5) -- (4.5, 2);
\draw[line width=0.5mm] (6, 2.5) -- (6, 2);
\draw[line width=0.5mm] (7.5, 2.5) -- (7.5, 2);
\draw[line width=0.5mm] (0.5, 1.5) -- (1, 1.5); \draw[color=black!30!green, very thick](1.5, 1.5) circle (0.5); \draw (1.5, 1.5) node (X) {$G_B$}; \draw[color=black!30!green, line width=0.5mm] (1.9, 1.2) -- (2.4, 0.9);
\draw[line width=0.5mm] (2, 1.5) -- (2.5, 1.5); \draw[color=black!30!green, very thick](3, 1.5) circle (0.5); \draw (3, 1.5) node {$G_B$}; \draw[color=black!30!green, line width=0.5mm] (3.4, 1.2) -- (3.9, 0.9);
\draw[line width=0.5mm] (3.5, 1.5) -- (4, 1.5); \draw[color=black!30!green, very thick](4.5, 1.5) circle (0.5); \draw (4.5, 1.5) node {$G_B$}; \draw[color=black!30!green, line width=0.5mm] (4.9, 1.2) -- (5.4, 0.9);
\draw[line width=0.5mm] (5, 1.5) -- (5.5, 1.5); \draw[color=black!30!green, very thick](6, 1.5) circle (0.5); \draw (6, 1.5) node {$G_B$}; \draw[color=black!30!green, line width=0.5mm] (6.4, 1.2) -- (6.9, 0.9);
\draw[line width=0.5mm] (6.5, 1.5) -- (7, 1.5); \draw[color=black!30!green, very thick](7.5, 1.5) circle (0.5); \draw (7.5, 1.5) node {$G_B$}; \draw[color=black!30!green, line width=0.5mm] (7.9, 1.2) -- (8.4, 0.9);
\draw[line width=0.5mm] (8, 1.5) -- (8.5, 1.5);
\draw[line width=0.5mm] (1.5, 1) -- (1.5, 0.5);
\draw[line width=0.5mm] (3, 1) -- (3, 0.5);
\draw[line width=0.5mm] (4.5, 1) -- (4.5, 0.5);
\draw[line width=0.5mm] (6, 1) -- (6, 0.5);
\draw[line width=0.5mm] (7.5, 1) -- (7.5, 0.5);

\draw[line width=0.5mm] (0.5, 0) -- (1, 0); \draw[color=black!30!green, very thick](1.5, 0) circle (0.5); \draw (1.5, 0) node (X) {$G_B$}; \draw[color=black!30!green, line width=0.5mm] (1.9, -0.3) -- (2.4, -0.6);
\draw[line width=0.5mm] (2, 0) -- (2.5, 0); \draw[color=black!30!green, very thick](3, 0) circle (0.5); \draw (3, 0) node {$G_B$}; \draw[color=black!30!green, line width=0.5mm] (3.4, -0.3) -- (3.9, -0.6);
\draw[line width=0.5mm] (3.5, 0) -- (4, 0); \draw[color=black!30!green, very thick](4.5, 0) circle (0.5); \draw (4.5, 0) node {$G_B$}; \draw[color=black!30!green, line width=0.5mm] (4.9, -0.3) -- (5.4, -0.6);
\draw[line width=0.5mm] (5, 0) -- (5.5, 0); \draw[color=black!30!green, very thick](6, 0) circle (0.5); \draw (6, 0) node {$G_B$}; \draw[color=black!30!green, line width=0.5mm] (6.4, -0.3) -- (6.9, -0.6);
\draw[line width=0.5mm] (6.5, 0) -- (7, 0); \draw[color=black!30!green, very thick](7.5, 0) circle (0.5); \draw (7.5, 0) node {$G_B$}; \draw[color=black!30!green, line width=0.5mm] (7.9, -0.3) -- (8.4, -0.6);
\draw[line width=0.5mm] (8, 0) -- (8.5, 0);
\draw[line width=0.5mm] (1.5, -0.5) -- (1.5, -1);
\draw[line width=0.5mm] (3, -0.5) -- (3, -1);
\draw[line width=0.5mm] (4.5, -0.5) -- (4.5, -1);
\draw[line width=0.5mm] (6, -0.5) -- (6, -1);
\draw[line width=0.5mm] (7.5, -0.5) -- (7.5, -1);

\draw[line width=0.5mm] (0.5, -1.5) -- (1, -1.5); \draw[color=black!30!green, very thick](1.5, -1.5) circle (0.5); \draw (1.5, -1.5) node (X) {$G_B$}; \draw[color=black!30!green, line width=0.5mm] (1.9, -1.8) -- (2.4, -2.1);
\draw[line width=0.5mm] (2, -1.5) -- (2.5, -1.5); \draw[color=black!30!green, very thick](3, -1.5) circle (0.5); \draw (3, -1.5) node {$G_B$}; \draw[color=black!30!green, line width=0.5mm] (3.4, -1.8) -- (3.9, -2.1);
\draw[line width=0.5mm] (3.5, -1.5) -- (4, -1.5); \draw[color=black!30!green, very thick](4.5, -1.5) circle (0.5); \draw (4.5, -1.5) node {$G_B$}; \draw[color=black!30!green, line width=0.5mm] (4.9, -1.8) -- (5.4, -2.1);
\draw[line width=0.5mm] (5, -1.5) -- (5.5, -1.5); \draw[color=black!30!green, very thick](6, -1.5) circle (0.5); \draw (6, -1.5) node {$G_B$}; \draw[color=black!30!green, line width=0.5mm] (6.4, -1.8) -- (6.9, -2.1);
\draw[line width=0.5mm] (6.5, -1.5) -- (7, -1.5); \draw[color=black!30!green, very thick](7.5, -1.5) circle (0.5); \draw (7.5, -1.5) node {$G_B$}; \draw[color=black!30!green, line width=0.5mm] (7.9, -1.8) -- (8.4, -2.1);
\draw[line width=0.5mm] (8, -1.5) -- (8.5, -1.5);
\draw[line width=0.5mm] (1.5, -2) -- (1.5, -2.5);
\draw[line width=0.5mm] (3, -2) -- (3, -2.5);
\draw[line width=0.5mm] (4.5, -2) -- (4.5, -2.5);
\draw[line width=0.5mm] (6, -2) -- (6, -2.5);
\draw[line width=0.5mm] (7.5, -2) -- (7.5, -2.5);
\draw (0, -1.5) node {$\ldots$};
\draw (9, -1.5) node {$\ldots$};
\draw (4.5, 3.2) node {$\vdots$};
\draw (4.5, -5.8) node {$\vdots$};

\draw[line width=0.5mm] (0.5, -3) -- (1, -3); \draw[color=black!30!green, very thick](1.5, -3) circle (0.5); \draw (1.5, -3) node {$G_B$}; \draw[color=black!30!green, line width=0.5mm] (1.9, -3.3) -- (2.4, -3.6);
\draw[line width=0.5mm] (2, -3) -- (2.5, -3); \draw[color=black!30!green, very thick](3, -3) circle (0.5); \draw (3, -3) node {$G_B$}; \draw[color=black!30!green, line width=0.5mm] (3.4, -3.3) -- (3.9, -3.6);
\draw[line width=0.5mm] (3.5, -3) -- (4, -3); \draw[color=black!30!green, very thick](4.5, -3) circle (0.5); \draw (4.5, -3) node {$G_B$}; \draw[color=black!30!green, line width=0.5mm] (4.9, -3.3) -- (5.4, -3.6);
\draw[line width=0.5mm] (5, -3) -- (5.5, -3); \draw[color=black!30!green, very thick](6, -3) circle (0.5); \draw (6, -3) node {$G_B$}; \draw[color=black!30!green, line width=0.5mm] (6.4, -3.3) -- (6.9, -3.6);
\draw[line width=0.5mm] (6.5, -3) -- (7, -3); \draw[color=black!30!green, very thick](7.5, -3) circle (0.5); \draw (7.5, -3) node {$G_B$}; \draw[color=black!30!green, line width=0.5mm] (7.9, -3.3) -- (8.4, -3.6);
\draw[line width=0.5mm] (8, -3) -- (8.5, -3);
\draw[line width=0.5mm] (1.5, -3.5) -- (1.5, -4);
\draw[line width=0.5mm] (3, -3.5) -- (3, -4);
\draw[line width=0.5mm] (4.5, -3.5) -- (4.5, -4);
\draw[line width=0.5mm] (6, -3.5) -- (6, -4);
\draw[line width=0.5mm] (7.5, -3.5) -- (7.5, -4);

\draw[line width=0.5mm] (0.5, -4.5) -- (1, -4.5); \draw[color=black!30!green, very thick](1.5, -4.5) circle (0.5); \draw (1.5, -4.5) node {$G_B$}; \draw[color=black!30!green, line width=0.5mm] (1.9, -4.8) -- (2.4, -5.1);
\draw[line width=0.5mm] (2, -4.5) -- (2.5, -4.5); \draw[color=black!30!green, very thick](3, -4.5) circle (0.5); \draw (3, -4.5) node {$G_B$}; \draw[color=black!30!green, line width=0.5mm] (3.4, -4.8) -- (3.9, -5.1);
\draw[line width=0.5mm] (3.5, -4.5) -- (4, -4.5); \draw[color=black!30!green, very thick](4.5, -4.5) circle (0.5); \draw (4.5, -4.5) node {$G_B$}; \draw[color=black!30!green, line width=0.5mm] (4.9, -4.8) -- (5.4, -5.1);
\draw[line width=0.5mm] (5, -4.5) -- (5.5, -4.5); \draw[color=black!30!green, very thick](6, -4.5) circle (0.5); \draw (6, -4.5) node {$G_B$}; \draw[color=black!30!green, line width=0.5mm] (6.4, -4.8) -- (6.9, -5.1);
\draw[line width=0.5mm] (6.5, -4.5) -- (7, -4.5); \draw[color=black!30!green, very thick](7.5, -4.5) circle (0.5); \draw (7.5, -4.5) node {$G_B$}; \draw[color=black!30!green, line width=0.5mm] (7.9, -4.8) -- (8.4, -5.1);
\draw[line width=0.5mm] (8, -4.5) -- (8.5, -4.5);
\draw[line width=0.5mm] (1.5, -5) -- (1.5, -5.5);
\draw[line width=0.5mm] (3, -5) -- (3, -5.5);
\draw[line width=0.5mm] (4.5, -5) -- (4.5, -5.5);
\draw[line width=0.5mm] (6, -5) -- (6, -5.5);
\draw[line width=0.5mm] (7.5, -5) -- (7.5, -5.5);
\end{diagram},
\label{eq:genfuncGPhi}
\end{equation}
where $G_B$ stands for the generating function of the impurity tensor and takes the following position-dependent form:
\begin{equation}
\begin{diagram}
\draw[line width=0.5mm] (1.1, -0.3) -- (1.5, -0.9);
\draw[line width=0.5mm] (0.5, -1.3) -- (1.3, -1.3); \draw[color=green!40!gray, very thick](1.8, -1.3) circle (0.5); \draw (1.8, -1.3) node {$G_B$};
\draw[line width=0.5mm] (2.3, -1.3) -- (3.1, -1.3);
\draw[line width=0.5mm] (2.1, -1.7) -- (2.5, -2.3);
\draw[color=green!40!gray, line width=0.5mm] (1.8, -1.8) -- (1.8, -2.5);
\end{diagram}
=
\begin{diagram}
\draw[line width=0.5mm] (1.1, -0.3) -- (1.5, -0.9);
\draw[line width=0.5mm] (0.5, -1.3) -- (1.3, -1.3); \draw[color=red!60, very thick](1.8, -1.3) circle (0.5); \draw (1.8, -1.3) node {$A$};
\draw[line width=0.5mm] (2.3, -1.3) -- (3.1, -1.3);
\draw[line width=0.5mm] (2.1, -1.7) -- (2.5, -2.3);
\draw[color=red!60, line width=0.5mm] (1.8, -1.8) -- (1.8, -2.5);
\end{diagram}
+\lambda e^{-i\textbf{k}\cdot\textbf{r}}
\begin{diagram}
\draw[line width=0.5mm] (1.1, -0.3) -- (1.5, -0.9);
\draw[line width=0.5mm] (0.5, -1.3) -- (1.3, -1.3); \draw[color=blue!60, very thick](1.8, -1.3) circle (0.5); \draw (1.8, -1.3) node {$B$};
\draw[line width=0.5mm] (2.3, -1.3) -- (3.1, -1.3);
\draw[line width=0.5mm] (2.1, -1.7) -- (2.5, -2.3);
\draw[color=blue!60, line width=0.5mm] (1.8, -1.8) -- (1.8, -2.5);
\end{diagram} \;.
\label{eq:genefuncB}
\end{equation}
The one-particle excited state in Eq.~\eqref{eq:excitation} can now be evaluated by taking the derivative:
\begin{equation}
    |\Phi_\textbf{k}(B)\rangle = \frac{\partial }{\partial \lambda}|G_{\Phi}(\lambda)\rangle\Bigr |_{\lambda=0}.
\end{equation}

One question immediately follows: namely, what would be the right environment tensors for $\langle G_\Phi(\lambda)|G_\Phi(\lambda)\rangle$ when computing physical observables? Note that, for PEPS in the thermodynamic limit, the contraction scheme relies on the translation invariance to interpret the environment tensors as fixed points of the one-dimensional transfer operator. For the generating function in Eq.~\eqref{eq:genfuncGPhi}, this translation invariance is lost for any nonzero $\lambda$. The solution to this issue comes from the following fact: after taking derivatives at $\lambda=0$, all contributions of the sum can be viewed as correlation functions of local operators. Recall that when computing observables or correlation functions in the infinite MPS, one uses the fixed point of the transfer matrix as the boundary, suggesting that fixed-point tensors of the transfer matrix of the infinite MPS ground state are the right environment tensors for generating function of infinite MPS. The same is true for the PEPS ground state in the thermodynamic limit. Indeed, we have tested that using a fixed point of the MPS transfer matrix in the generating functions gives the same result as directly summing all tensor diagrams. Thus, similarly for PEPS, we can use the environment tensors from CTMRG of the ground state PEPS (i.e., the state $|G_{\Phi}(\lambda)\rangle$ at $\lambda=0$) as the environment tensors for the generating function Eq.~\eqref{eq:genfuncGPhi}.

With the generating function for the excited state, the norm matrix and effective Hamiltonian can be similarly expressed as derivatives of a single network. Using translation symmetry, we can lower the order of derivatives by introducing two slightly new networks~\cite{Tu2021}. For that, we first construct two (idealized) generating functions, $G_{\mathbf{N}}(\lambda, B)$ and $G_{\mathbf{H}}(\lambda, \mu, B)$ for norm matrix and effective Hamiltonian, respectively. $G_{\mathbf{N}}$ takes the following form:
\begin{equation}
G_{\mathbf{N}}(\lambda, B) = 
\begin{diagram}
\draw (-1, 2) .. controls (-1, 2) and (-1, 2.5) .. (-0.5, 2.5);
\draw (7, 2) .. controls (7, 2) and (7, 2.5) .. (6.5, 2.5);
\draw (7, -5) .. controls (7, -5) and (7, -5.5) .. (6.5, -5.5);
\draw (-1, -5) .. controls (-1, -5) and (-1, -5.5) .. (-0.5, -5.5);
\draw[dashed] (-0.5, 2.5) -- (6.5, 2.5);
\draw[dashed] (7, 2) -- (7, -5);
\draw[dashed] (-1, 2) -- (-1, -5);
\draw[dashed] (-0.5, -5.5) -- (6.5, -5.5);

\draw[color=black!60, very thick](-2.5, 4) rectangle (-1.5, 3); \draw (-2, 3.5) node {$C_1$};
\draw[color=black!30!green, line width=1mm] (-1.5, 3.5) -- (-1, 3.5);
\draw[color=black!30!green, line width=1mm] (-2, 3) -- (-2, 2.5);
\draw[color=black!60, very thick](-1, 4.2) rectangle (0, 3); \draw (-0.5, 3.5) node {$T_1$};
\draw[color=black!30!green, line width=1mm] (0, 3.5) -- (0.5, 3.5);
\draw[line width=1mm] (-0.5, 3) -- (-0.5, 2.5);
\draw[color=black!60, very thick](-2.7, 2.5) rectangle (-1.5, 1.5); \draw (-2, 2) node {$T_4$};
\draw[color=black!30!green, line width=1mm] (-2, 1.5) -- (-2, 1);
\draw[line width=1mm] (-1.5, 2) -- (-1, 2);
\draw (1, 3.5) node {$\ldots$};
\draw (-2, 0.7) node {$\vdots$};

\draw[color=black!60, very thick](8.5, 4) rectangle (7.5, 3); \draw (8, 3.5) node {$C_2$};
\draw[color=black!30!green, line width=1mm] (7, 3.5) -- (7.5, 3.5); 
\draw[color=black!30!green, line width=1mm] (8, 3) -- (8, 2.5);
\draw[color=black!60, very thick](8.7, 2.5) rectangle (7.5, 1.5); \draw (8, 2) node {$T_2$};
\draw[color=black!30!green, line width=1mm] (8, 1.5) -- (8, 1);
\draw[line width=1mm] (7, 2) -- (7.5, 2);
\draw[color=black!60, very thick](6, 4.2) rectangle (7, 3); \draw (6.5, 3.5) node {$T_1$};
\draw[color=black!30!green, line width=1mm] (5.5, 3.5) -- (6, 3.5);
\draw[line width=1mm] (6.5, 3) -- (6.5, 2.5);
\draw (5, 3.5) node {$\ldots$};
\draw (8, 0.7) node {$\vdots$};

\draw[color=black!60, very thick](8.5, -6) rectangle (7.5, -7); \draw (8, -6.5) node {$C_3$};
\draw[color=black!30!green, line width=1mm] (7, -6.5) -- (7.5, -6.5); 
\draw[color=black!30!green, line width=1mm] (8, -5.5) -- (8, -6);
\draw[color=black!60, very thick](8.7, -4.5) rectangle (7.5, -5.5); \draw (8, -5) node {$T_2$};
\draw[color=black!30!green, line width=1mm] (8, -4) -- (8, -4.5);
\draw[line width=1mm] (7, -5) -- (7.5, -5);
\draw[color=black!60, very thick](6, -6) rectangle (7, -7.2); \draw (6.5, -6.5) node {$T_3$};
\draw[color=black!30!green, line width=1mm] (5.5, -6.5) -- (6, -6.5);
\draw[line width=1mm] (6.5, -5.5) -- (6.5, -6);
\draw (5, -6.5) node {$\ldots$};
\draw (8, -3.3) node {$\vdots$};

\draw[color=black!60, very thick](-2.5, -6) rectangle (-1.5, -7); \draw (-2, -6.5) node {$C_4$};
\draw[color=black!30!green, line width=1mm] (-1.5, -6.5) -- (-1, -6.5); 
\draw[color=black!30!green, line width=1mm] (-2, -5.5) -- (-2, -6);
\draw[color=black!60, very thick](-1, -6) rectangle (0, -7.2); \draw (-0.5, -6.5) node {$T_3$};
\draw[color=black!30!green, line width=1mm] (0, -6.5) -- (0.5, -6.5);
\draw[line width=1mm] (-0.5, -5.5) -- (-0.5, -6);
\draw[color=black!60, very thick](-2.7, -4.5) rectangle (-1.5, -5.5); \draw (-2, -5) node {$T_4$};
\draw[color=black!30!green, line width=1mm] (-2, -4) -- (-2, -4.5);
\draw[line width=1mm] (-1.5, -5) -- (-1, -5);
\draw (1, -6.5) node {$\ldots$};
\draw (-2, -3.3) node {$\vdots$};

\draw[line width=1mm] (1.5, 1) -- (1.5, 0.5);
\draw[line width=1mm] (3, 1) -- (3, 0.5);
\draw[line width=1mm] (4.5, 1) -- (4.5, 0.5);
\draw[line width=1mm] (0.5, 0) -- (1, 0); \fill[color=green!40!gray, very thick](1.5, 0) circle (0.5);
\draw[line width=1mm] (2, 0) -- (2.5, 0); \fill[color=green!40!gray, very thick](3, 0) circle (0.5);
\draw[line width=1mm] (3.5, 0) -- (4, 0); \fill[color=green!40!gray, very thick](4.5, 0) circle (0.5);
\draw[line width=1mm] (5, 0) -- (5.5, 0);
\draw[line width=1mm] (1.5, -0.5) -- (1.5, -1);
\draw[line width=1mm] (3, -0.5) -- (3, -1);
\draw[line width=1mm] (4.5, -0.5) -- (4.5, -1);

\draw[line width=1mm] (0.5, -1.5) -- (1, -1.5); \fill[color=green!40!gray, very thick](1.5, -1.5) circle (0.5); \draw (1.5, -1.5) node (X) {};
\draw[line width=1mm] (2, -1.5) -- (2.5, -1.5); \draw[color=green!40!gray, very thick](3, -1.5) circle (0.5); \draw[color=green!40!gray, line width=0.8mm] (2.6, -1.1) -- (3.4, -1.9); \draw[color=green!40!gray, line width=0.8mm] (2.6, -1.9) -- (3.4, -1.1);
\draw[line width=1mm] (3.5, -1.5) -- (4, -1.5); \fill[color=green!40!gray, very thick](4.5, -1.5) circle (0.5);
\draw[line width=1mm] (5, -1.5) -- (5.5, -1.5);
\draw[line width=1mm] (1.5, -2) -- (1.5, -2.5);
\draw[line width=1mm] (3, -2) -- (3, -2.5);
\draw[line width=1mm] (4.5, -2) -- (4.5, -2.5);

\draw[line width=1mm] (0.5, -3) -- (1, -3); \fill[color=green!40!gray, very thick](1.5, -3) circle (0.5);
\draw[line width=1mm] (2, -3) -- (2.5, -3); \fill[color=green!40!gray, very thick](3, -3) circle (0.5);
\draw[line width=1mm] (3.5, -3) -- (4, -3); \fill[color=green!40!gray, very thick](4.5, -3) circle (0.5);
\draw[line width=1mm] (5, -3) -- (5.5, -3);
\draw[line width=1mm] (1.5, -3.5) -- (1.5, -4);
\draw[line width=1mm] (3, -3.5) -- (3, -4);
\draw[line width=1mm] (4.5, -3.5) -- (4.5, -4);
\draw (0, -1.5) node {$\ldots$};
\draw (6, -1.5) node {$\ldots$};
\draw (3, 1.7) node {$\vdots$};
\draw (3, -4.3) node {$\vdots$};
\end{diagram}.
\label{eq:genfuncNorm}
\end{equation}
The green tensors in Eq.~\eqref{eq:genfuncNorm} are given by:
\begin{equation}
\begin{diagram}
\draw[line width=1mm] (1.5, -0.3) -- (1.5, -0.8);
\draw[line width=1mm] (0.5, -1.3) -- (1, -1.3); \fill[color=green!40!gray, very thick](1.5, -1.3) circle (0.5);
\draw[line width=1mm] (2, -1.3) -- (2.5, -1.3);
\draw[line width=1mm] (1.5, -1.8) -- (1.5, -2.3);
\end{diagram}
=
\begin{diagram}
\draw[line width=0.5mm] (1.1, 1.0) -- (1.5, 0.4);
\draw[line width=0.5mm] (0.5, 0) -- (1.3, 0); \draw[color=green!40!gray, very thick](1.8, 0) circle (0.5); \draw (1.8, 0) node {$G_B$};
\draw[line width=0.5mm] (2.3, 0) -- (3.1, 0);
\draw[line width=0.5mm] (2.1, -0.4) -- (2.5, -1);
\draw[color=green!40!gray, line width=0.5mm] (1.8, -0.5) -- (1.8, -1.25);
\draw[color=red!60, line width=0.5mm] (1.8, -1.25) -- (1.8, -2);
\draw[line width=0.5mm] (1.1, -1.5) -- (1.5, -2.1);
\draw[line width=0.5mm] (0.5, -2.5) -- (1.3, -2.5); \draw[color=red!60, very thick](1.8, -2.5) circle (0.5); \draw (1.8, -2.5) node {$\bar{A}$};
\draw[line width=0.5mm] (2.3, -2.5) -- (3.1, -2.5);
\draw[line width=0.5mm] (2.1, -2.9) -- (2.5, -3.5);
\end{diagram}
\;\;\;\;\text{and}\;\;\;\;
\begin{diagram}
\draw[line width=1mm] (1.5, -0.3) -- (1.5, -0.8);
\draw[line width=1mm] (0.5, -1.3) -- (1, -1.3); \draw[color=green!40!gray, very thick](1.5, -1.3) circle (0.5); \draw[color=green!40!gray, line width=0.8mm] (1.1, -0.9) -- (1.9, -1.7); \draw[color=green!40!gray, line width=0.8mm] (1.1, -1.7) -- (1.9, -0.9);
\draw[line width=1mm] (2, -1.3) -- (2.5, -1.3);
\draw[line width=1mm] (1.5, -1.8) -- (1.5, -2.3);
\end{diagram}
=
\begin{diagram}
\draw[line width=0.5mm] (1.1, -0.3) -- (1.5, -0.9);
\draw[line width=0.5mm] (0.5, -1.3) -- (1.3, -1.3); \draw[color=green!40!gray, very thick](1.8, -1.3) circle (0.5); \draw (1.8, -1.3) node {$G_B$};
\draw[line width=0.5mm] (2.3, -1.3) -- (3.1, -1.3);
\draw[line width=0.5mm] (2.1, -1.7) -- (2.5, -2.3);
\draw[color=green!40!gray, line width=0.5mm] (1.8, -1.8) -- (1.8, -2.5);
\end{diagram},
\label{eq:phystraceG}
\end{equation}
where the tensor with a cross mark lies in the center of the lattice. The  norm matrix can then be evaluated by taking the derivative:
\begin{equation}
\textbf{N} = \frac{\partial }{\partial B}G_{\mathbf{N}}(\lambda, B)\Bigr |_{\lambda=1, B=0}.
\label{eq:normMat}
\end{equation}

For the generating function of the effective Hamiltonian, we need to insert the local Hamiltonian between the bra and the ket layers. Note that, although the local Hamiltonian can be represented as a projected entangled-pair operator, it is rarely used in practice, due to the increased computational cost. Here we use the generating function for a local Hamiltonian, as introduced in Ref.~\cite{Tu2021}, to represent the full Hamiltonian as a derivative of a new network. Then we sandwich the generating function of the local Hamiltonian term between the bra and ket layers. To illustrate, assuming that we only have the nearest neighbor terms, a new tensor can be defined as:
\begin{equation}
\begin{diagram}
\draw[line width=1mm] (1.5, -0.3) -- (1.5, -0.8);
\draw[line width=1mm] (0.5, -1.3) -- (1, -1.3); 
\fill[color=green!40!gray, very thick](1.5, -1.3) circle (0.5);
\draw[color=red!60, line width=1mm] (2, -1.3) -- (2.5, -1.3);
\draw[line width=1mm] (1.5, -1.8) -- (1.5, -2.3);
\end{diagram}
=
\begin{diagram}
\draw[line width=0.5mm] (1.1, 1.0) -- (1.5, 0.4);
\draw[line width=0.5mm] (0.5, 0) -- (1.3, 0); \draw[color=green!40!gray, very thick](1.8, 0) circle (0.5); \draw (1.8, 0) node {$G_B$};
\draw[line width=0.5mm] (2.3, 0) -- (3.1, 0);
\draw[line width=0.5mm] (2.1, -0.4) -- (2.5, -1);
\draw[color=green!40!gray, line width=0.5mm] (1.8, -0.5) -- (1.8, -1);

\draw[color=red!60, line width=0.5mm] (1.8, -2) -- (1.8, -2.5);
\draw[line width=0.5mm] (1.1, -2) -- (1.5, -2.6);
\draw[line width=0.5mm] (0.5, -3) -- (1.3, -3); \draw[color=red!60, very thick](1.8, -3) circle (0.5); \draw (1.8, -3) node {$\bar{A}$};
\draw[line width=0.5mm] (2.3, -3) -- (3.1, -3);
\draw[line width=0.5mm] (2.1, -3.4) -- (2.5, -4);

\clip (1.5, -1) rectangle (3.1, -2);
\filldraw[rounded corners, fill={rgb, 255: red, 255; green, 38; blue, 0}] (1.5, -1) rectangle (4.5, -2); \draw (2.5, -1.5) node {$G_{\hat{H}}$};
\end{diagram}
\;\;\;\;\text{and}\;\;\;\;
G_{\hat{H}}=\mathbb{I}+ \mu \hat{H}_{\textbf{r}, \textbf{r}+\hat{i}},
\label{eq:redbond}
\end{equation}
where $\mathbb{I}$ is the identity matrix and $\hat{i}=\hat{x}, \hat{y}$.
Then the (idealized) generating function for the effective Hamiltonian is given by the following:
\begin{equation}
G_{\mathbf{H}}(\lambda, \mu, B) = 
\begin{diagram}
\draw (-1, 2) .. controls (-1, 2) and (-1, 2.5) .. (-0.5, 2.5);
\draw (7, 2) .. controls (7, 2) and (7, 2.5) .. (6.5, 2.5);
\draw (7, -5) .. controls (7, -5) and (7, -5.5) .. (6.5, -5.5);
\draw (-1, -5) .. controls (-1, -5) and (-1, -5.5) .. (-0.5, -5.5);
\draw[dashed] (-0.5, 2.5) -- (6.5, 2.5);
\draw[dashed] (7, 2) -- (7, -5);
\draw[dashed] (-1, 2) -- (-1, -5);
\draw[dashed] (-0.5, -5.5) -- (6.5, -5.5);

\draw[color=black!60, very thick](-2.5, 4) rectangle (-1.5, 3); \draw (-2, 3.5) node {$C_1$};
\draw[color=black!30!green, line width=1mm] (-1.5, 3.5) -- (-1, 3.5);
\draw[color=black!30!green, line width=1mm] (-2, 3) -- (-2, 2.5);
\draw[color=black!60, very thick](-1, 4.2) rectangle (0, 3); \draw (-0.5, 3.5) node {$T_1$};
\draw[color=black!30!green, line width=1mm] (0, 3.5) -- (0.5, 3.5);
\draw[line width=1mm] (-0.5, 3) -- (-0.5, 2.5);
\draw[color=black!60, very thick](-2.7, 2.5) rectangle (-1.5, 1.5); \draw (-2, 2) node {$T_4$};
\draw[color=black!30!green, line width=1mm] (-2, 1.5) -- (-2, 1);
\draw[line width=1mm] (-1.5, 2) -- (-1, 2);
\draw (1, 3.5) node {$\ldots$};
\draw (-2, 0.7) node {$\vdots$};

\draw[color=black!60, very thick](8.5, 4) rectangle (7.5, 3); \draw (8, 3.5) node {$C_2$};
\draw[color=black!30!green, line width=1mm] (7, 3.5) -- (7.5, 3.5); 
\draw[color=black!30!green, line width=1mm] (8, 3) -- (8, 2.5);
\draw[color=black!60, very thick](8.7, 2.5) rectangle (7.5, 1.5); \draw (8, 2) node {$T_2$};
\draw[color=black!30!green, line width=1mm] (8, 1.5) -- (8, 1);
\draw[line width=1mm] (7, 2) -- (7.5, 2);
\draw[color=black!60, very thick](6, 4.2) rectangle (7, 3); \draw (6.5, 3.5) node {$T_1$};
\draw[color=black!30!green, line width=1mm] (5.5, 3.5) -- (6, 3.5);
\draw[line width=1mm] (6.5, 3) -- (6.5, 2.5);
\draw (5, 3.5) node {$\ldots$};
\draw (8, 0.7) node {$\vdots$};

\draw[color=black!60, very thick](8.5, -6) rectangle (7.5, -7); \draw (8, -6.5) node {$C_3$};
\draw[color=black!30!green, line width=1mm] (7, -6.5) -- (7.5, -6.5); 
\draw[color=black!30!green, line width=1mm] (8, -5.5) -- (8, -6);
\draw[color=black!60, very thick](8.7, -4.5) rectangle (7.5, -5.5); \draw (8, -5) node {$T_2$};
\draw[color=black!30!green, line width=1mm] (8, -4) -- (8, -4.5);
\draw[line width=1mm] (7, -5) -- (7.5, -5);
\draw[color=black!60, very thick](6, -6) rectangle (7, -7.2); \draw (6.5, -6.5) node {$T_3$};
\draw[color=black!30!green, line width=1mm] (5.5, -6.5) -- (6, -6.5);
\draw[line width=1mm] (6.5, -5.5) -- (6.5, -6);
\draw (5, -6.5) node {$\ldots$};
\draw (8, -3.3) node {$\vdots$};

\draw[color=black!60, very thick](-2.5, -6) rectangle (-1.5, -7); \draw (-2, -6.5) node {$C_4$};
\draw[color=black!30!green, line width=1mm] (-1.5, -6.5) -- (-1, -6.5); 
\draw[color=black!30!green, line width=1mm] (-2, -5.5) -- (-2, -6);
\draw[color=black!60, very thick](-1, -6) rectangle (0, -7.2); \draw (-0.5, -6.5) node {$T_3$};
\draw[color=black!30!green, line width=1mm] (0, -6.5) -- (0.5, -6.5);
\draw[line width=1mm] (-0.5, -5.5) -- (-0.5, -6);
\draw[color=black!60, very thick](-2.7, -4.5) rectangle (-1.5, -5.5); \draw (-2, -5) node {$T_4$};
\draw[color=black!30!green, line width=1mm] (-2, -4) -- (-2, -4.5);
\draw[line width=1mm] (-1.5, -5) -- (-1, -5);
\draw (1, -6.5) node {$\ldots$};
\draw (-2, -3.3) node {$\vdots$};

\draw[color=red!60, line width=1mm] (1.5, 1) -- (1.5, 0.5);
\draw[color=red!60, line width=1mm] (3, 1) -- (3, 0.5);
\draw[color=red!60, line width=1mm] (4.5, 1) -- (4.5, 0.5);
\draw[color=red!60, line width=1mm] (0.5, 0) -- (1, 0); \fill[color=green!40!gray, very thick](1.5, 0) circle (0.5);
\draw[color=red!60, line width=1mm] (2, 0) -- (2.5, 0); \fill[color=green!40!gray, very thick](3, 0) circle (0.5);
\draw[color=red!60, line width=1mm] (3.5, 0) -- (4, 0); \fill[color=green!40!gray, very thick](4.5, 0) circle (0.5);
\draw[color=red!60, line width=1mm] (5, 0) -- (5.5, 0);
\draw[color=red!60, line width=1mm] (1.5, -0.5) -- (1.5, -1);
\draw[color=red!60, line width=1mm] (3, -0.5) -- (3, -1);
\draw[color=red!60, line width=1mm] (4.5, -0.5) -- (4.5, -1);

\draw[color=red!60, line width=1mm] (0.5, -1.5) -- (1, -1.5); \fill[color=green!40!gray, very thick](1.5, -1.5) circle (0.5); \draw (1.5, -1.5) node (X) {};
\draw[color=red!60, line width=1mm] (2, -1.5) -- (2.5, -1.5); \draw[color=green!40!gray, very thick](3, -1.5) circle (0.5); \draw[color=green!40!gray, line width=0.8mm] (2.6, -1.1) -- (3.4, -1.9); \draw[color=green!40!gray, line width=0.8mm] (2.6, -1.9) -- (3.4, -1.1);
\draw[color=red!60, line width=1mm] (3.5, -1.5) -- (4, -1.5); \fill[color=green!40!gray, very thick](4.5, -1.5) circle (0.5);
\draw[color=red!60, line width=1mm] (5, -1.5) -- (5.5, -1.5);
\draw[color=red!60, line width=1mm] (1.5, -2) -- (1.5, -2.5);
\draw[color=red!60, line width=1mm] (3, -2) -- (3, -2.5);
\draw[color=red!60, line width=1mm] (4.5, -2) -- (4.5, -2.5);

\draw[color=red!60, line width=1mm] (0.5, -3) -- (1, -3); \fill[color=green!40!gray, very thick](1.5, -3) circle (0.5);
\draw[color=red!60, line width=1mm] (2, -3) -- (2.5, -3); \fill[color=green!40!gray, very thick](3, -3) circle (0.5);
\draw[color=red!60, line width=1mm] (3.5, -3) -- (4, -3); \fill[color=green!40!gray, very thick](4.5, -3) circle (0.5);
\draw[color=red!60, line width=1mm] (5, -3) -- (5.5, -3);
\draw[color=red!60, line width=1mm] (1.5, -3.5) -- (1.5, -4);
\draw[color=red!60, line width=1mm] (3, -3.5) -- (3, -4);
\draw[color=red!60, line width=1mm] (4.5, -3.5) -- (4.5, -4);
\draw (0, -1.5) node {$\ldots$};
\draw (6, -1.5) node {$\ldots$};
\draw (3, 1.7) node {$\vdots$};
\draw (3, -4.3) node {$\vdots$};
\end{diagram}.
\label{eq:genfuncHam}
\end{equation}
From Eq.~\eqref{eq:genfuncHam}, the effective Hamiltonian $\textbf{H}$ can be obtained by taking a second order derivative 
\begin{equation} 
\label{eq:hamMat}
\textbf{H} = \frac{\partial^2 }{\partial B\partial\mu}G_{\mathbf{H}}(\lambda, \mu, B)\Bigr |_{\lambda=1, \mu=0, B=0} \;.
\end{equation}

Denoting the linear size of the bulk, i.e., the region encircled by the dashed box in Eqs.~\eqref{eq:genfuncNorm} and \eqref{eq:genfuncHam}, as $L_x$ and $L_y$, it appears that $L_x$ and $L_y$ have to be chosen carefully. In the case of infinite MPS, we have checked that the linear size in the bulk has to be compatible with the momenta, i.e., $k=2\pi m/L_x$ ($m=0,1,2,\cdots, L_x-1$), to ensure exact zero modes in the norm matrix. We expect that the same requirement is also true in the two-dimensional case. Moreover, $L_x$ and $L_y$ should be large enough~(compared to the maximal correlation length of the ground state) to mimic an infinite system, so that the error due to finite linear size becomes negligible. At the same time, with larger linear sizes, more $\textbf{k}$ points can be considered. Note that, here we restrict to a finite bulk size, instead of an infinite one, due to the fact that for a given ground state tensor $A$ and impurity tensor $B$, the generating function and its derivatives may not converge within the same number of iterations. This is because the forward calculation is computing the zeroth order of the expansion of generating functions at $\lambda=0$, which typically resembles the tensor diagram for a local observable (see Eq.~\eqref{eq:genfuncNorm} and \eqref{eq:normMat} for example), while the derivatives contain multi-point correlators. Therefore, it is conservative but safe to work with a finite bulk size.

With the norm matrix and effective Hamiltonian obtained from derivatives of generating functions, we can now solve the generalized eigenvalue equation in Eq.~\eqref{eq:eigenvalue}, and obtain the impurity tensor $\mathbf{B}$ for excited states. As mentioned before, due to gauge degree of freedom, exact zero modes appear in the norm matrix and one needs to project out the corresponding subspace, leading to the following generalized eigenvalue equation: 
\begin{equation}
(P^{\dagger}\mathbf{H}P)_{\mu\nu}\mathbf{B_{\nu}} = E(P^{\dagger}\mathbf{N}P)_{\mu\nu}\mathbf{B}_{\nu},
\label{eq:eigenvalue_projected}
\end{equation}
where the subspace projector $P$ projects out the exact zero modes of the norm matrix. Note that, up to this point, there is no approximation involved and the scheme can be directly applied to the infinite MPS context where computation can be made essentially exact.

\begin{figure*}[!hbt]
    \centering
    \includegraphics[width=2.0\columnwidth]{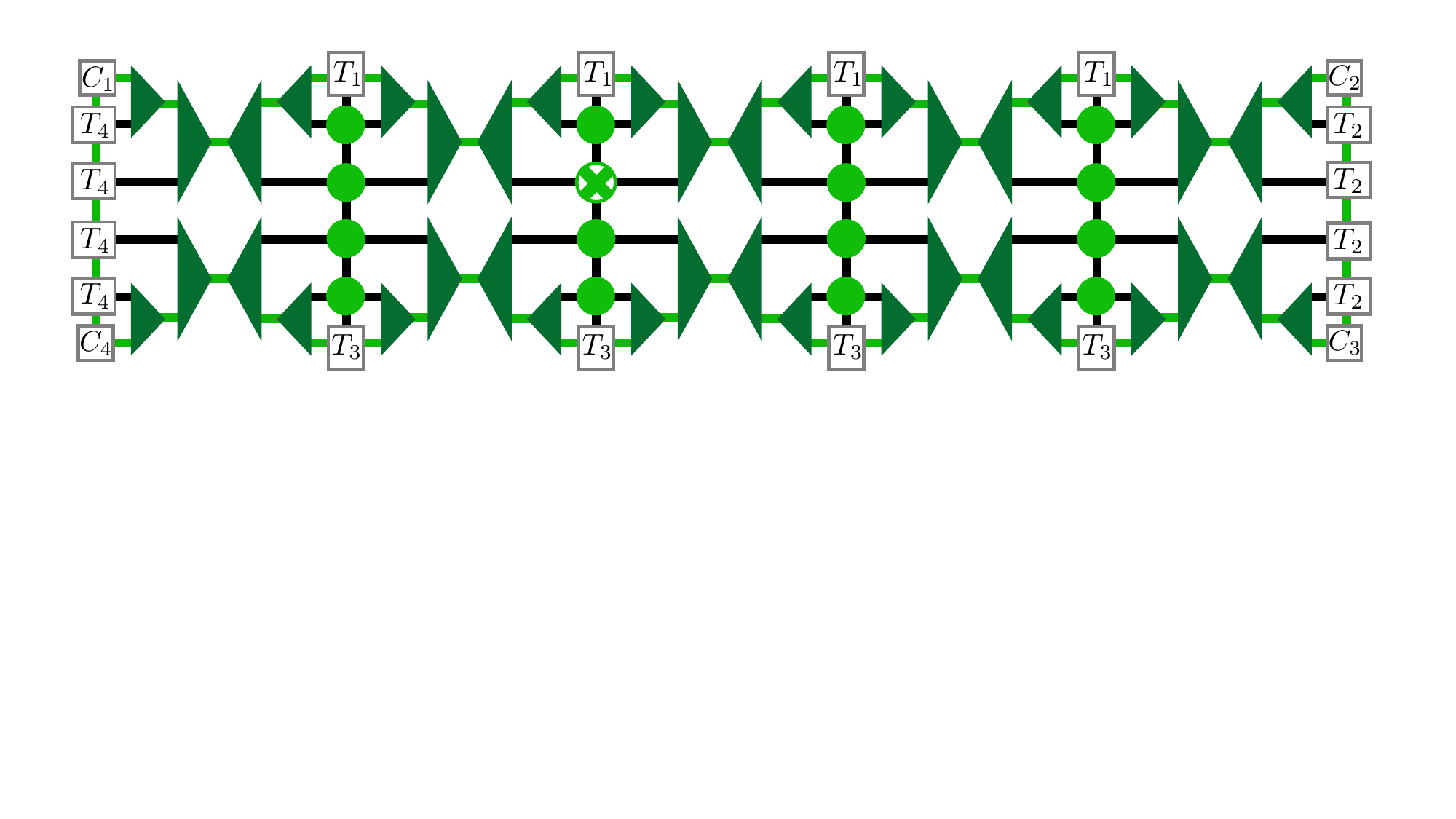}
    \caption{Complete tensor network for generating function of norm matrix with bulk size $N_s=4\times4$. The projectors inserted are taken from the CTMRG of the ground state. Similarly for the generating function of the effective Hamiltonian.}
    \label{fig:fig1}
\end{figure*}

However, there is one more issue for PEPS to address: for large tensor network graphs in 2D~(Eqs.~\eqref{eq:genfuncNorm} and \eqref{eq:genfuncHam}), its computational complexity grows exponentially with system size. Therefore, the exact contraction is not feasible. By utilizing the projectors during the CTMRG process, on the other hand, we can approximate tensors on each column as the following:
\begin{equation}
\begin{diagram}
\draw[color=black!30!green, line width=1mm] (2, 0) -- (2.5, 0); \draw[color=black!60, very thick](3.5, 0.7) rectangle (2.5, -0.5); \draw (3, 0) node {$T_1$};
\draw[color=black!30!green, line width=1mm] (3.5, 0) -- (4, 0);
\draw[line width=1mm] (3, -0.5) -- (3, -1);
\draw[color=black!60, very thick](5, 0.7) rectangle (4, -0.5); \draw (4.5, 0) node {$T_1$};
\draw[line width=1mm] (4.5, -0.5) -- (4.5, -1);
\draw[color=black!30!green, line width=1mm] (5.5, 0) -- (5, 0);

\draw[line width=1mm] (2, -1.5) -- (2.5, -1.5); \fill[color=black!30!green, very thick](3, -1.5) circle (0.5); \draw[line width=1mm] (3.5, -1.5) -- (4, -1.5);
\draw[line width=1mm] (3, -2) -- (3, -2.5);
\fill[color=black!30!green, very thick](4.5, -1.5) circle (0.5);
\draw[line width=1mm] (5.5, -1.5) -- (5, -1.5);
\draw[line width=1mm] (4.5, -2) -- (4.5, -2.5);

\draw[line width=1mm] (2, -3) -- (2.5, -3); \fill[color=black!30!green, very thick](3, -3) circle (0.5); \draw[line width=1mm] (3.5, -3) -- (4, -3);
\draw[line width=1mm] (3, -3.5) -- (3, -4);
\fill[color=black!30!green, very thick](4.5, -3) circle (0.5);
\draw[line width=1mm] (5.5, -3) -- (5, -3);
\draw[line width=1mm] (4.5, -3.5) -- (4.5, -4);
\draw (3, -4.2) node {$\vdots$};
\draw (4.5, -4.2) node {$\vdots$};
\end{diagram}
\approx
\begin{diagram}
\filldraw[color=black!60!green] (8.5,-0.75) -- (9.5,0.25) -- (9.5,-1.75) -- cycle; \draw[color=black!30!green, line width=1mm] (8, -0.75) -- (8.5, -0.75);
\filldraw[color=black!60!green] (5,-0.75) -- (4,0.25) -- (4,-1.75) -- cycle; \draw[color=black!30!green, line width=1mm] (5, -0.75) -- (5.5, -0.75);

\filldraw[color=black!60!green] (7,-1.75) -- (8,0) -- (8,-3.5) -- cycle; 
\filldraw[color=black!60!green] (6.5,-1.75) -- (5.5,0) -- (5.5,-3.5) -- cycle; \draw[color=black!30!green, line width=1mm] (6.5, -1.75) -- (7, -1.75);

\draw[color=black!30!green, line width=1mm] (2, 0) -- (2.5, 0); \draw[color=black!60, very thick](3.5, 0.7) rectangle (2.5, -0.5); \draw (3, 0) node {$T_1$};
\draw[color=black!30!green, line width=1mm] (3.5, 0) -- (4, 0);
\draw[line width=1mm] (3, -0.5) -- (3, -1);

\draw[line width=1mm] (2, -1.5) -- (2.5, -1.5); \fill[color=black!30!green, very thick](3, -1.5) circle (0.5); \draw[line width=1mm] (3.5, -1.5) -- (4, -1.5);
\draw[line width=1mm] (3, -2) -- (3, -2.5);

\draw[line width=1mm] (2, -3) -- (2.5, -3); \fill[color=black!30!green, very thick](3, -3) circle (0.5); \draw[line width=1mm] (3.5, -3) -- (5.5, -3);
\draw[line width=1mm] (3, -3.5) -- (3, -4);
\draw (3, -4.2) node {$\vdots$};

\draw[color=black!30!green, line width=1mm] (9.5, 0) -- (10, 0);
\draw[color=black!60, very thick](11, 0.7) rectangle (10, -0.5); \draw (10.5, 0) node {$T_1$};
\draw[color=black!30!green, line width=1mm] (11, 0) -- (11.5, 0);
\draw[line width=1mm] (10.5, -0.5) -- (10.5, -1);

\draw[line width=1mm] (9.5, -1.5) -- (10, -1.5); \fill[color=black!30!green, very thick](10.5, -1.5) circle (0.5); \draw[line width=1mm] (11, -1.5) -- (11.5, -1.5);
\draw[line width=1mm] (10.5, -2) -- (10.5, -2.5);

\draw[line width=1mm] (8, -3) -- (10, -3); \fill[color=black!30!green, very thick](10.5, -3) circle (0.5); \draw[line width=1mm] (11, -3) -- (11.5, -3);
\draw[line width=1mm] (10.5, -3.5) -- (10.5, -4);
\draw (10.5, -4.2) node {$\vdots$};
\end{diagram},
\label{eq:projector}
\end{equation}
where the tensor projectors~(dark green triangles) can be evaluated through the singular value decomposition~\cite{Corboz2014} and a similar process can be conducted for the row and corner tensors. With the projectors inserted, now we show the complete tensor network for the norm matrix in Fig.~\ref{fig:fig1}. Note that here we show explicitly how to compress our tensor graphs vertically but in fact the similar action is also taken along the horizontal direction. We can then take the average of these two compressed tensor graphs as $G_{\mathbf{N}}(\lambda, B)$.

The motivation for the above mentioned splitting the contraction of $G_{\mathbf{N}}(\lambda, B)$ into two is due to the insertion of local Hamiltonian terms for $G_{\mathbf{H}}(\lambda, \mu, B)$. Note that the tensor projectors intuitively would reduce the correlation between nearby sites, which is undesirable when inserting the local Hamiltonian operators. Therefore, in practice, we conduct the insertion of $\hat{H}_{\textbf{r}, \textbf{r}+\hat{x}}$ and $\hat{H}_{\textbf{r}, \textbf{r}+\hat{y}}$ separately, resulting in the tensor contractions along the horizontal and vertical directions, respectively. More specifically, we include $G_{\hat{H}}$ in all the vertical~(horizontal) bonds and compress the tensor graph for each column~(row).

The scheme is now clear and all we need to do is to first contract the {\it finite} tensor network in Eqs.~\eqref{eq:genfuncNorm} and \eqref{eq:genfuncHam} (with tensor projectors inserted), and then compute the derivatives using either AD or any other ways of numerical finite difference approach. Then, by solving the eigenvalue equation Eq.~\eqref{eq:eigenvalue_projected} after getting $\textbf{N}$ and $\textbf{H}$, we obtain the impurity tensors with corresponding energies. With zero modes in the norm matrix projected out, the excited states constructed with the impurity tensors are then orthogonal to each other, and also orthogonal to the ground state.

Note that, due to the projectors inserted in Fig.~\ref{fig:fig1}, the zero modes in the norm matrix are no longer exact, and instead become fuzzy. Therefore, we use the subspace projector $P$ to truncate the basis by discarding the norm matrix eigenvector with eigenvalue close to zero~\cite{Ponsioen2020}. Denoting the eigenvalue decomposition of $\textbf{N}$ as $\textbf{N}=v\Lambda v^{\dagger}$, we keep the desired subspace of $v$ and construct $\tilde{v}$, which serves as $P$ in Eq.~(\ref{eq:eigenvalue_projected}). One subtle point is that the selection of subspace cannot be simply determined a priori. Practically, we start from the leading eigenvalue of norm matrix and gradually enlarge the size of the subspace by including next leading ones. If adding certain norm matrix eigenvector changes the energy eigenvalues drastically, we would discard this vector when constructing the subspace. So far, we cannot conclude a strict protocol in choosing the subspace projectors, and occasionally need to examine different threshold for the subspace to obtain stable energy eigenvalues. Note also that, a suitable gauge in the ground state tensor may be helpful for improving the conditioning of the norm matrix~\cite{Lubasch2014}. With these measures, we solve Eq.~(\ref{eq:eigenvalue_projected}), and the excited state can then be constructed with $\textbf{B}_{\nu}$ for further investigation of its properties.

\subsection{Discussion of the algorithm}
\label{subsec:diss_algorithm}

Comparing to the previous generating functions for the finite-size MPS~\cite{Tu2021}, two new ingredients are included in the infinite PEPS case. The first one is concerning the infinite size in typical PEPS calculations. Although our scheme can be easily generalized to the finite PEPS, due to the large computational cost, iPEPS is typically used in the literature. For that purpose, we have introduced ground state environment tensors in computing the norm matrix and effective Hamiltonian of the variational space. 

The second ingredient is about the contraction of the generating functions, where we have used ground state projectors to achieve an efficient contraction. The idea of using projectors from the ground state CTMRG procedure can further be put on firm grounds as follows. That is, all the correlations in our ansatz are essentially mediated by the ground state tensors, and all contributions to the effective Hamiltonian and norm matrix can be viewed as correlation functions of local operators in the ground state. Unraveling the CTMRG procedure, one would find that the CTMRG procedure is essentially inserting tensor projectors~(which was computed self-consistently) into the original network. Moreover, when conducting the forward calculation of generating functions with AD, we assign the impurity tensor being a tensor with all elements equal to zero (see Eqs.~\eqref{eq:normMat} and \eqref{eq:hamMat} where the derivatives are computed at $B=0$). Therefore, the contraction is akin to the normal ground state CTMRG process. Thus it is natural to use the same projectors for the generating functions, which would not lower the accuracy. However, if one computes the generating function for excited state at nonzero $\lambda$ or nonzero $B$ tensor, which would indeed be the case when part of the derivatives is obtained with finite difference approaches, the projectors may need to be recomputed, taking into account the effect of $B$ tensor. In that case, the computation time and memory consumption would also be higher. Nevertheless, here we refrain from making a detailed comparison between various possibilities, but leave them to future developments. We also note that this scheme is not restricted to CTMRG and in fact can be generalized to other PEPS contraction methods, e.g., tensor renormalization group~\cite{Xie2012} and boundary MPS approach~\cite{Jordan2008}.

In comparison with previous approaches using PEPS for constructing one-particle excited states, our approach largely reduces the number of tensor diagrams to be considered. In Ref.~\onlinecite{Vanderstraeten2015} the tensor contraction is done by evaluating the corner-shaped transfer matrix, with norm matrix and effective Hamiltonian obtained by summing up the tensor diagrams whose impurity tensors in the bra and ket layer are located in different relative positions. While the corner transfer matrices are also adopted in Ref.~\onlinecite{Ponsioen2020} for the environment tensors, again due to the position change many tensor graphs need to be taken into account.

The approach proposed in Ref.~\onlinecite{Ponsioen2022} comes closer to the approach presented here, which relies on AD to do the tensor diagram summations. Let us now compare with that one here. The most prominent feature of our approach is that we only need to deal with one tensor diagram for $\textbf{N}$ or $\textbf{H}$, where all the tensors are fixed before contraction. In Ref.~\onlinecite{Ponsioen2022}, one needs to take care of the additional environment tensors~(coming from impurity tensors), which were computed iteratively. In this sense, the approach presented here gets rid of all the intricate summations and updating steps in PEPS excitations. This simplification will play an essential role when considering multi-particle excitations with PEPS.

Before showing the applications, let us also mention that the generating functions are in fact independent of AD. In some cases, the derivative is taken only with respect to one parameter $\lambda$ and therefore can be easily carried out using the finite difference method. Indeed, using finite difference may become crucial when the bond dimension is large, since in that case AD might be limited due to memory issues. Nevertheless, in this work we use AD for two reasons. Firstly, in obtaining the norm matrix and effective Hamiltonian, multiple parameters are involved and thus the back propagation mode of AD is more efficient. Secondly, being numerically exact, AD can avoid finite difference error. While the full expression of norm matrix and effective Hamiltonian may not be necessary for Lanczos type diagonalization algorithms (and thus only one parameter is involved in the derivative), we will not explore this possibility here and leave it to future considerations.

\begin{figure*}
\centering
	\includegraphics[width=2.0\columnwidth]{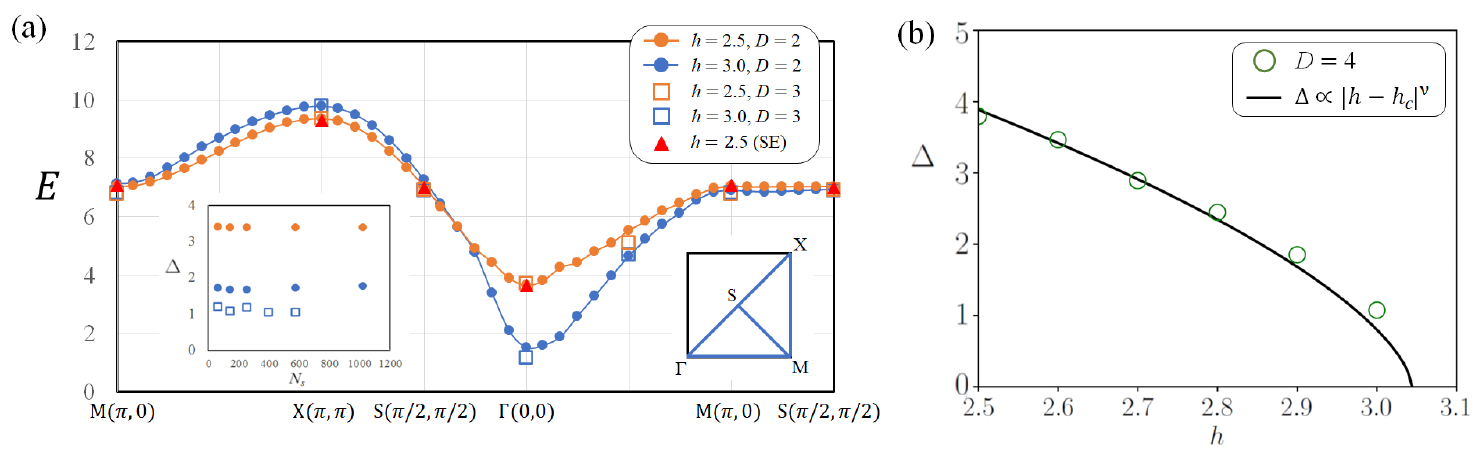}
\caption{(a) The energy dispersion of the lowest-lying excitation for transverse-field Ising model at distinct $\textbf{k}$ points for $h=2.5$ and $h=3.0$. We adopt $N_s=24\times24$ for $D=2$ and $N_s=16\times16$ for $D=3$ and confirm that further enlarging $N_s$ does not change the results. The red triangles show the results from series expansion~(SE). The left inset shows the energy gap with different $N_s$ for $(h,D,\xi)=(2.5,2,0.603)$, $(h,D,\xi)=(3.0,2,1.194)$, and $(h,D,\xi)=(3.0,3,2.531)$, where $\xi$ is the maximal correlation length. The right inset shows the high symmetry path in the Brillouin zone. (b) The gap function along with magnetic field $h$. We adopt $D=4, N_s=12 \times 12$, and the solid curve takes the scaling relation of 3D Ising universality class with the critical exponent $\nu=0.629971$.}
\label{fig:fig2}
\end{figure*}

\section{Applications}
\label{sec:application}

Now we benchmark our algorithm with two well-studied models, namely the transverse-field Ising model and spin-$1/2$ Heisenberg model on the square lattice, and then move to the more challenging $J_1-J_2$ model.

\subsection{Transverse-field Ising model}
\label{subsec:TFI}

The spin-1/2 quantum Ising model with a transverse field in 2D is described by:
\begin{equation}
H_{\text{TFI}}=-\sum_{\langle i,j\rangle}\sigma^z_i\sigma^z_j-h\sum_i\sigma^x_i,
\label{eq:TFI}
\end{equation}
where $\sigma^{x,z}$ are the Pauli matrices and the summation of $\langle i,j\rangle$ runs over all nearest-neighbor pairs. As the most widely known model for benchmarking, its phase transition can be accurately captured by PEPS~\cite{Jordan2008, Chen2022}. A high precision estimate of the transition point was made by the cluster Monte Carlo method with the value being $h_c=3.04438$~\cite{Blote2002}. When $h\to\infty$, the ground state preserves the global $\mathbb{Z}_2$ symmetry $\hat{U} = \prod_i \sigma^x_i$ with a non-zero energy gap, while at $h=0$ the ground state is two-fold degenerate with all spins aligning up or down, breaking the $\mathbb{Z}_2$ symmetry. Close to the transition point, the energy gap decreases and becomes zero at $h_c$, which is a Lorentz-invariant critical point.

In order to compare with the results in Refs.~\onlinecite{Vanderstraeten2019a} and \onlinecite{Ponsioen2020}, we first compute the lowest-energy excited state at each $\textbf{k}$ and plot its energy versus momentum in Fig.~\ref{fig:fig2}(a). Here and below we have chosen the high symmetry path $M\to X \to S \to \Gamma \to M \to S$ in the first Brilliouin zone of the square lattice, shown in the right inset of Fig.~\ref{fig:fig2}(a).

Let us first examine the choice of bulk size $N_s=L_x\times L_y$ in computing excited states. As shown in the left inset of Fig.~\ref{fig:fig2}(a), with $h$ closer to the critical point and larger bond dimension $D$, the maximal correlation length ($\xi$) of the ground state (measured from transfer matrix spectrum of ground state PEPS) becomes larger, and one would need larger bulk size $N_s$ to converge the energy gap $\Delta$ (at momenta $\Gamma (0,0)$), in agreement with previous discussion on bulk size. Nevertheless, due to the relatively small bond dimension (and therefore small $\xi$) used in this work, the excited states shown in this work are relatively easy to converge.

From Fig.~\ref{fig:fig2}(a), one can further find that away from the critical point (e.g., at $h=2.5$), the finite $D$ effect is rather small, and the excitation energy does not show significant changes from $D=2$ to $D=3$. Close to the critical point (e.g., at $h=3.0$), the finite $D$ effect is most prominent in the lowest excited state, where the energy gap $\Delta$ decreases with larger bond dimension. Our results show both qualitative and quantitative agreement with Ref.~\onlinecite{Vanderstraeten2019a} and \onlinecite{Ponsioen2020}, and also agree with the series expansion result~\cite{book.Oitmaa}.

Besides the quasiparticle dispersion relation, we can also fit the numerically computed energy gap $\Delta$ versus magnetic field $h$, taking a function form $\Delta\propto |h-h_c|^\nu$. In Fig.~\ref{fig:fig2}(b), we compare the numerical data with the analytically known critical exponent $\nu=0.629971$, and indeed find a good agreement, further confirming the accuracy of the generating function method.

\subsection{Heisenberg model}
\label{subsec:Heisenberg}

Next, we examine another model widely used for the purpose of benchmarking, the antiferromagnetic (AFM) Heisenberg model:
\begin{equation}
H_{\text{Heisenberg}}=J\sum_{\langle i,j\rangle}\mathbf{S}_i\cdot\mathbf{S}_j,
\label{eq:Heisenberg}
\end{equation}
where we take $J=1$ as the energy unit, and the sum runs over all nearest-neighbor pairs. While the AFM nature demands at least a two-site unit cell for the ground state ansatz, by rotating the local spin on one of the two sublattices, we can preserve the one-site translation symmetry for PEPS.

\begin{figure}
\centering
	\includegraphics[width=1.0\columnwidth]{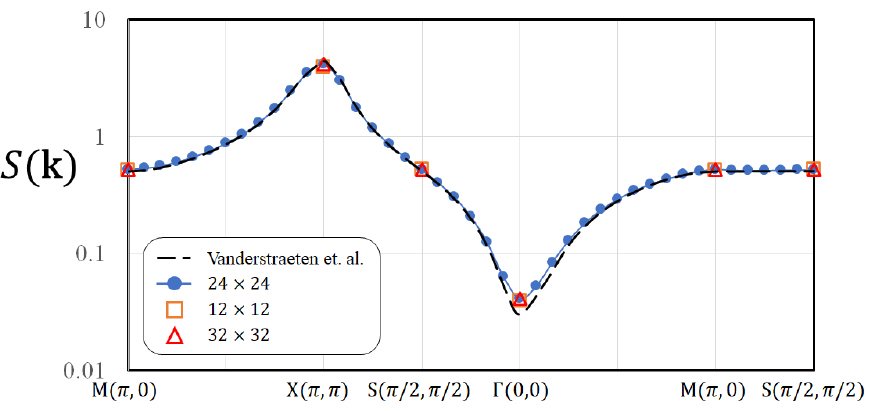}
\caption{Static structure factor $S(\textbf{k})$ for the Heisenberg model~($D=4$, $\xi=1.8438$), with three different $N_s$. For comparison, the black dashed line shows the results of Ref.~\cite{Vanderstraeten2016} with $D=4$.}
\label{fig:fig3}
\end{figure}

Before showing the excited states, as a good benchmark we compute the static structure factor, which can be measured by elastic neutron scattering experiments. The static structure factor is defined by the following equation~\cite{Zheng2005}:
\begin{equation}
S(\textbf{k})=\sum_{\textbf{r},\alpha}e^{i\textbf{k}\cdot\textbf{r}}\langle S^\alpha_0 S^\alpha_\textbf{r} \rangle_c =\sum_\alpha S^\alpha(\textbf{k}),
\label{eq:stat}
\end{equation}
where $\langle \cdot \rangle_c$ denotes the connected part of the correlator and $\alpha=x,y,z$. Because of the AFM nature, the static structure factor possesses a peak at $\textbf{k}=(\pi,\pi)$, and becomes zero at $\textbf{k}=(0,0)$. Given a variational PEPS ground state, the static structure factor can be evaluated efficiently using generating functions. Here we show $S(\textbf{k})$ with $D=4$ in Fig.~\ref{fig:fig3}. Our result is in good agreement with the previous one using iPEPS~\cite{Vanderstraeten2016}, as well as those by other methods~\cite{Zheng2005, Luscher2009}. More importantly, the results of different $N_s$ all look similar, fortifying the fact that for our method the choice of bulk size casts little effect on the final outcomes, as long as the bulk linear size is large comparing to the ground state maximal correlation length $\xi$.

\begin{figure}
\centering
	\includegraphics[width=0.95\columnwidth]{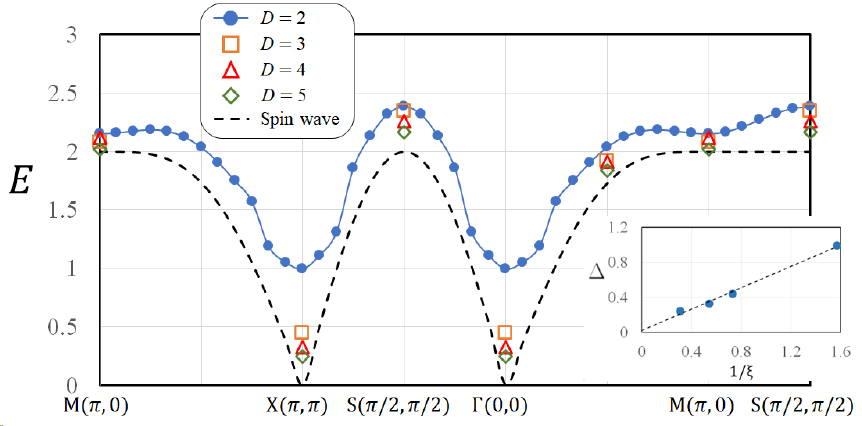}
\caption{Energy dispersion of lowest-lying excitations for Heisenberg model with $D=2$~($N_s=24\times24$), $D=3$~($N_s=16\times16$), $D=4$~($N_s=12\times12$), and $D=5$~($N_s=8\times8$). The dashed line indicates the result from the spin wave theory. The inset shows the scaling of $\Delta$ (at $\Gamma (0,0)$) with the inverse of maximal correlation length $1/\xi$. Using a linear fit, $\Delta$ extrapolates to a value close to zero with $\xi\to \infty$.}
\label{fig:fig4}
\end{figure}

In Fig.~\ref{fig:fig4} we show our results for the lowest-energy dispersion of the Heisenberg model. One clear feature that we can see is the vanishing of excitation energy at $\Gamma(0,0)$ and $\text{X}(\pi,\pi)$ with increasing $D$, which is desirable since the Heisenberg model is known to be gapless at the these two points. To further confirm the gapless nature, in the inset we provide the scaling of energy gap $\Delta$ (at $\Gamma(0,0)$) with $1/\xi$. We can see that the energy gap indeed extrapolates to a value close to zero when $\xi\to\infty$. Except for the gapped feature due to the finite bond dimension, our dispersion is close to the ones by earlier iPEPS results~\cite{Vanderstraeten2019a, Ponsioen2020} and Gutzwiller-projected trial wavefunctions~\cite{Piazza2015}. Also, the well-known feature that the gap at $(\pi,0)$ is smaller than the one at $(\pi/2,\pi/2)$~\cite{Piazza2015} is recovered in our simulation. In fact, the nature around $\text{M}(\pi,0)$ point is of interest and has been under investigation \cite{Shao2017, Verresen2018, Powalski2018}. In Ref.~\onlinecite{Ponsioen2020} it has been demonstrated using PEPS that the repulsion from the multi-magnon branches pushes the magnon at $\text{M}(\pi,0)$ to a lower energy~\cite{Verresen2018}. This feature is also captured in our simulation by showing a dip in the single magnon branch near the $\text{M}$ point, which cannot be captured by the spin wave theory~(dashed line). In the next section we will show that after introducing the next-nearest-neighbor (NNN) $J_2$ coupling the lowest excitation energy at $\text{M}(\pi,0)$ can further decrease.

\subsection{Spin-\texorpdfstring{$1/2$}{1/2} \texorpdfstring{$J_1$}{J1}-\texorpdfstring{$J_2$}{J2} antiferromagnet}
\label{subsec:J1J2}

\begin{figure}
\centering
	\includegraphics[width=0.99\columnwidth]{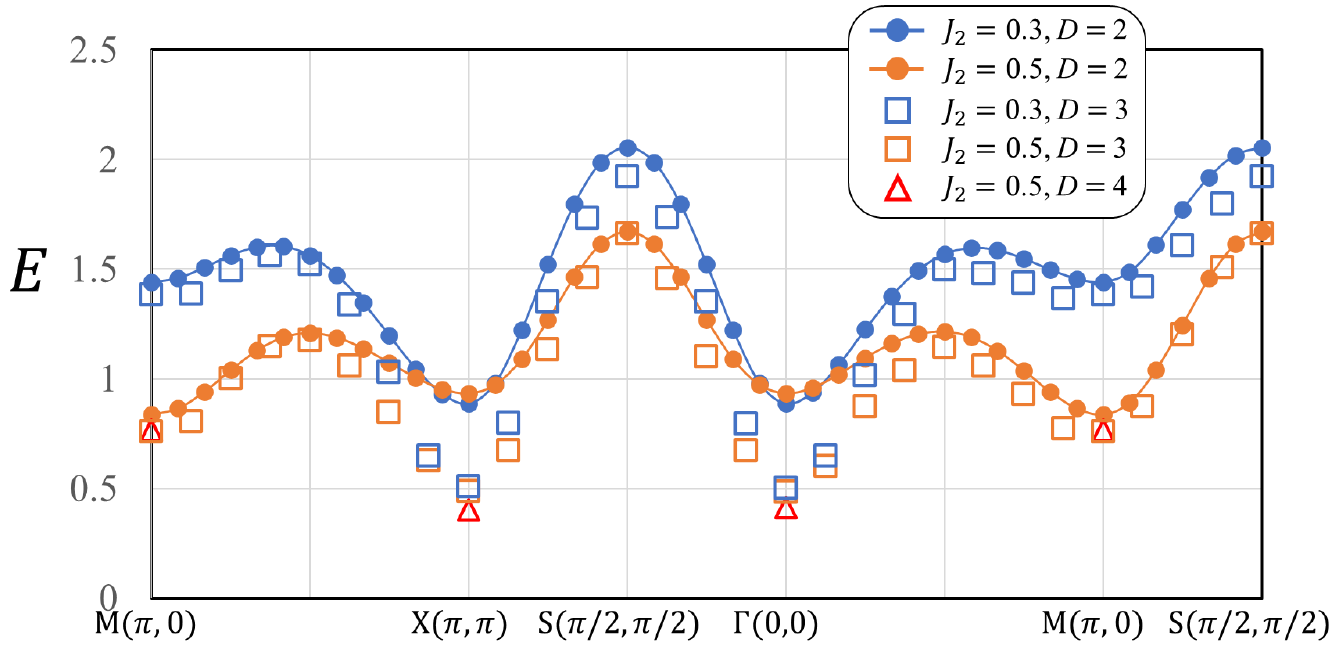}
\caption{The energy dispersion for the lowest excited state of the $J_1-J_2$ model at distinct $\textbf{k}$ points for $J_2=0.3$ and $J_2=0.5$. We adopt $N_s=24 \times 24$ for $D=2$, $N_s=16 \times 16$ for $D=3$, and $N_s=8 \times 8$ for $D=4$.}
\label{fig:fig5}
\end{figure}

\begin{figure*}
\centering
	\includegraphics[width=2\columnwidth]{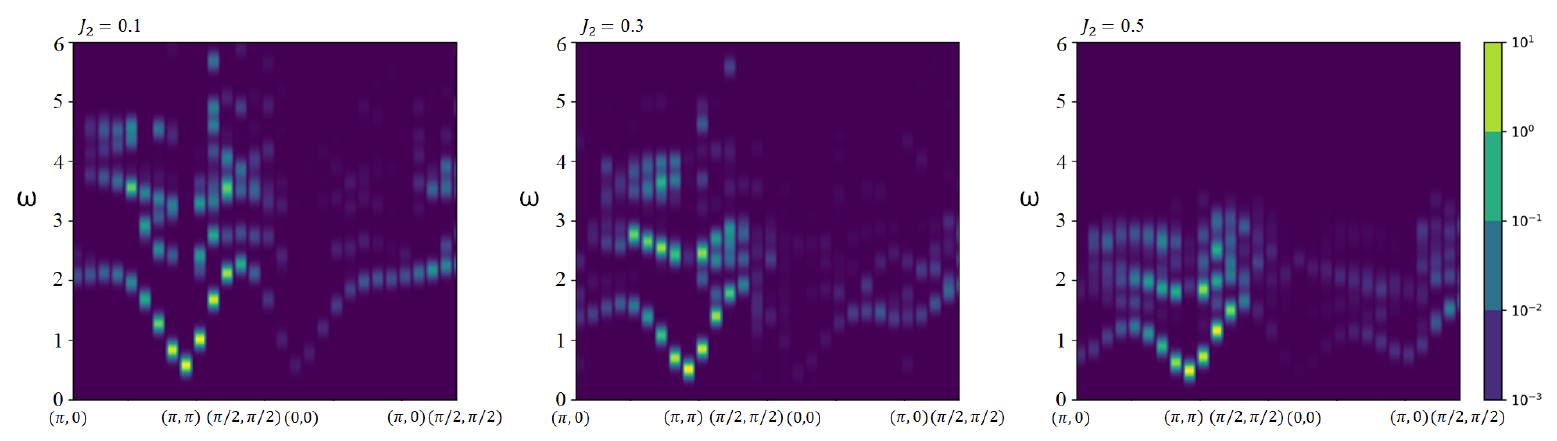}
\caption{The dynamical structure factor for $J_1$-$J_2$ model with $D=3$. Here we show results for $J_2=0.1$ (left panel), $J_2=0.3$ (middle panel), and $J_2=0.5$ (right panel), with $N_s=16 \times 16$. We replace the delta function with normalized Gaussian broadening with width $\sigma=0.1$.}
\label{fig:fig6}
\end{figure*}

In the previous subsections, we have studied models with only nearest neighbor couplings using our method. Both the transverse-field Ising model and Heisenberg model have served as good benchmarks and have been studied earlier using conventional iPEPS construction for excited states~\cite{Vanderstraeten2019a, Ponsioen2020, Ponsioen2022, Vanderstraeten2015}. Here, we aim to push further and study the $J_1-J_2$ model where the NNN coupling introduces frustration and thus exotic ground states~\cite{Morita2015,Liu2022}. The Hamiltonian is given by:
\begin{equation}
    H_{J_1-J_2}=J_1\sum_{\langle i,j\rangle}\mathbf{S}_i\cdot\mathbf{S}_j+
    J_2\sum_{\langle\langle k, l\rangle\rangle}\mathbf{S}_k\cdot\mathbf{S}_l,
    \label{eq:J1J2}
\end{equation}
where the $J_1$ ($J_2$) term runs over all (next-)nearest-neighbor pairs on the square lattice.

The $J_1-J_2$ model has been under intense investigation. Since the frustration induces a sign problem for quantum Monte Carlo method, alternative numerical methods have been applied, suggesting that between $J_2\approx0.5J_1$ and $J_2\approx0.6J_1$ a ground state without magnetic order is realized (hereafter we fix $J_1=1$). Using PEPS methods, this vanishing of the staggered magnetization has been confirmed \cite{Hasik2021} using advanced scaling techniques -- for any finite bond dimension, it appears that a variational PEPS ground state yields a non-zero magnetization. Despite an abundant amount of related studies for the ground state, very few results have been reported for the excited state and there is no result so far using the two-dimensional tensor network ansatz~\footnote{Prior to our work we have only noted related study by MPS on a six-leg cylinder for $J_1-J_2$ model~\cite{VanDamme2022}.}. We will now investigate the excitation spectrum for this frustrated model.

In order to include the NNN coupling, $\hat{i}$ in Eq.~(\ref{eq:redbond}) now represents $\hat{x}, \hat{y}, \hat{x}+\hat{y}$, and $\hat{x}-\hat{y}$. As a result, the tensor contraction that one needs to apply when preparing $G_{\textbf{H}}$ becomes slightly cumbersome. For example, the tensor graphs that we need to consider now involve two columns at the same time:
\begin{equation}
\begin{diagram}
\draw[color=black!30!green, line width=1mm] (2, 0) -- (2.5, 0); \draw[color=black!60, very thick](3.5, 0.7) rectangle (2.5, -0.5); \draw (3, 0) node {$T_1$};
\draw[color=black!30!green, line width=1mm] (3.5, 0) -- (4, 0);
\draw[line width=1mm] (3, -0.5) -- (3, -1);
\draw[color=black!60, very thick](5, 0.7) rectangle (4, -0.5); \draw (4.5, 0) node {$T_1$};
\draw[line width=1mm] (4.5, -0.5) -- (4.5, -1);
\draw[color=black!30!green, line width=1mm] (5.5, 0) -- (5, 0);

\draw[line width=1mm] (2, -1.5) -- (2.5, -1.5); \fill[color=black!30!green, very thick](3, -1.5) circle (0.5); \draw[color=red!60, line width=1mm] (3.5, -1.5) -- (4, -1.5);
\draw[color=red!60, line width=1mm] (3, -2) -- (3, -2.5);
\fill[color=black!30!green, very thick](4.5, -1.5) circle (0.5);
\draw[line width=1mm] (5.5, -1.5) -- (5, -1.5);
\draw[color=red!60, line width=1mm] (4.5, -2) -- (4.5, -2.5);
\draw[color=red!60, line width=1mm] (3.3, -1.8) -- (4.2, -2.6);
\draw[color=red!60, line width=1mm] (3.3, -2.6) -- (4.2, -1.8);

\draw[line width=1mm] (2, -3) -- (2.5, -3); \fill[color=black!30!green, very thick](3, -3) circle (0.5); \draw[color=red!60, line width=1mm] (3.5, -3) -- (4, -3);
\draw[color=red!60, line width=1mm] (3, -3.5) -- (3, -4);
\fill[color=black!30!green, very thick](4.5, -3) circle (0.5);
\draw[line width=1mm] (5.5, -3) -- (5, -3);
\draw[color=red!60, line width=1mm] (4.5, -3.5) -- (4.5, -4);
\draw[color=red!60, line width=1mm] (3.3, -3.3) -- (4.2, -4.1);
\draw[color=red!60, line width=1mm] (3.3, -4.1) -- (4.2, -3.3);
\draw (3, -4.2) node {$\vdots$};
\draw (4.5, -4.2) node {$\vdots$};

\draw[color=black!30!green, line width=1mm] (7, 0) -- (7.5, 0); \draw[color=black!60, very thick](8.5, 0.7) rectangle (7.5, -0.5); \draw (8, 0) node {$T_1$};
\draw[color=black!30!green, line width=1mm] (8.5, 0) -- (9, 0);
\draw[line width=1mm] (8, -0.5) -- (8, -1);
\draw[color=black!60, very thick](10, 0.7) rectangle (9, -0.5); \draw (9.5, 0) node {$T_1$};
\draw[line width=1mm] (9.5, -0.5) -- (9.5, -1);
\draw[color=black!30!green, line width=1mm] (10.5, 0) -- (10, 0);
\draw (1.25, -1.5) node {$\hdots$};
\draw (6.25, -1.5) node {$\hdots$};
\draw (11.25, -1.5) node {$\hdots$};

\draw[line width=1mm] (7, -1.5) -- (7.5, -1.5); \fill[color=black!30!green, very thick](8, -1.5) circle (0.5); \draw[color=red!60, line width=1mm] (8.5, -1.5) -- (9, -1.5);
\draw[color=red!60, line width=1mm] (8, -2) -- (8, -2.5);
\fill[color=black!30!green, very thick](9.5, -1.5) circle (0.5);
\draw[line width=1mm] (10.5, -1.5) -- (10, -1.5);
\draw[color=red!60, line width=1mm] (9.5, -2) -- (9.5, -2.5);
\draw[color=red!60, line width=1mm] (8.3, -1.8) -- (9.2, -2.6);
\draw[color=red!60, line width=1mm] (8.3, -2.6) -- (9.2, -1.8);

\draw[line width=1mm] (7, -3) -- (7.5, -3); \fill[color=black!30!green, very thick](8, -3) circle (0.5); \draw[color=red!60, line width=1mm] (8.5, -3) -- (9, -3);
\draw[color=red!60, line width=1mm] (8, -3.5) -- (8, -4);
\fill[color=black!30!green, very thick](9.5, -3) circle (0.5);
\draw[line width=1mm] (10.5, -3) -- (10, -3);
\draw[color=red!60, line width=1mm] (9.5, -3.5) -- (9.5, -4);
\draw[color=red!60, line width=1mm] (8.3, -3.3) -- (9.2, -4.1);
\draw[color=red!60, line width=1mm] (8.3, -4.1) -- (9.2, -3.3);
\draw (8, -4.2) node {$\vdots$};
\draw (9.5, -4.2) node {$\vdots$};
\end{diagram},
\label{eq:projector2}
\end{equation}
which consumes more memory when storing the computational graph for the back propagation. Therefore, our calculation in this work is restricted to $D=4$ for the $J_1-J_2$ model and we will leave further consideration on increasing the available bond dimension, e.g., using symmetries in tensor networks to reduce computational cost, to future work. We note in passing that, the NNN coupling may also be handled using diagonal channel environments in Ref.~\onlinecite{Vanderstraeten2016}, which however differs from the CTMRG environment used in this work.

In Fig.~\ref{fig:fig5} we show the lowest-energy quasiparticle dispersion for the $J_1-J_2$ model with $J_2=0.3$ and $J_2=0.5$ along a high symmetry path in the Brillouin zone. Similar to the case for the Heisenberg model, with increasing bond dimension, the energy gap becomes smaller, in agreement with the fact that the phase is gapless~\cite{Wang2018}. Moreover, these dispersions are also close to those by the variational Monte Carlo~(VMC) method~\cite{Ferrari2018}. Besides the shape of the dispersion, we can clearly see that the energy gap becomes smaller at $\text{M}(\pi,0)$ when $J_2$ increases. More importantly, at $J_2=0.5$ our results using $D=2$ indicate that even the $\text{M}(\pi,0)$ point may become gapless. This softening of the dispersion at the M point is suggestive of the formation of a spin liquid phase at intermediate values, where the mode at the M point would correspond to a two-spinon state \cite{Ferrari2018}. 

Based on the variational wavefunctions for the excited states, it is straightforward to compute their contributions to the dynamical structure factor (DSF).
The DSF is defined as 
$
S^{\alpha}(\textbf{k},\omega)=\sum_{n}|M^{\alpha}_\textbf{k}|^2\delta(\omega-E^{\textbf{k}}_n+E_0),
$
with $M^{\alpha}_{\textbf{k}}=|\langle\Phi_{\textbf{k}}(B_n)|S^{\alpha}_{\textbf{k}}|\Psi(A)\rangle|$. $E_0$ and $E^{\textbf{k}}_n$ are the ground-state and excited-state energies for $|\Psi(A)\rangle$ and $|\Phi_{\textbf{k}}(B_n)\rangle$, respectively. And $S^{\alpha}_{\textbf{k}}
$ is the Fourier transform of local spin operator with $\alpha=(x,y,z)$. Note that, in this case the forward mode AD could be computationally cheaper than reverse mode AD, since only a few parameters are involved in the derivative. In Fig.~\ref{fig:fig6}, we show the DSF for $J_2=0.1, 0.3, 0.5$, respectively, where all data are computed with PEPS bond dimension $D=3$. The delta function in DSF has been replaced by a normalized Gaussian with broadening width $\sigma=0.1$. Two clear features can be seen from our data: (1) the energy gap gets softened at the $\text{M}(\pi,0)$ point, and (2) the spectral weights get closer to the magnon branch~(lowest excited energies) in the frustrated region. We have noted that in the previous results by VMC the authors also observed a similar trend in their DSF results, where a clear energy continuum is formed after $J_2/J_1\approx0.45$. In Fig.~\ref{fig:fig6} we can clearly see that as $J_2$ goes to 0.5, various dispersions gather toward the lowest magnon branch, featuring the potential formation of the energy continuum.

\section{Conclusion}
\label{sec:conclusion}

In this work, we have extended the generating functions introduced in Ref.~\onlinecite{Tu2021} to PEPS. After obtaining the well approximated ground state through variational optimization, we solve the eigenvalue problem in Eq.~(\ref{eq:eigenvalue_projected}) where effective Hamiltonian and norm matrices are evaluated by taking the derivatives of their corresponding generating functions. We then utilize the eigenstates as the impurity tensors to construct the one-particle excited state and employ the transverse-field Ising model and the Heisenberg model as benchmarks. Our results are consistent with previous ones~\cite{Vanderstraeten2019a, Ponsioen2020} using PEPS that requires the serial summation on different parts of tensor graphs. We extend our consideration to the model with the NNN coupling and study the $J_1-J_2$ model. Despite the increasing computational cost which hinders the computation with larger $D$, our results already show good agreement with previous ones by VMC~\cite{Ferrari2018}.

Before closing, let us discuss how to generalize this idea to multi-particle excitations, especially those associated with spinons. It is known that spin liquids are characterized by fractionalized excitations, where the quasiparticles cannot appear alone but come in pairs, which are intrinsically two-particle excitations. In the previous study, spinon excitation (among other fractional excitation) has been treated in MPS calculation using good quantum numbers, which appears as a local excitation with a non-local string attached~\cite{Zauner2018, Vanderstraeten2020}. In the case of PEPS, fractionalized excitations can also be defined and their static properties, e.g., correlation functions and correlation lengths have been studied in Ref.~\onlinecite{Chen2020}. In that case, the local tensor of PEPS needs a certain gauge symmetry. Once constructed, the excited states containing spinon excitations can be obtained. To further study dispersion relation of spinon excitation, one may put one spinon infinitely far away or use a suitable boundary condition to reduce the problem into an effective one-particle problem. Generating functions can then be used for efficiently optimizing the one-particle problem as we have shown above.

In summary, our proposal of combining the generating function with tensor network ansatz helps reduce the complexity in obtaining the physical properties or observables involving the summation of tensor networks, especially in computing quasiparticle excitations. The present method provides not only the dispersion relation of the quasiparticles but also a systematic way of knowing what the relevant quasiparticle for a given system is even if it is not clear a priori. It may then be possible to discover completely new quasiparticles in 2D quantum many-body systems. From another perspective, with the generating function method, the PEPS representation for excited state now takes almost the same form as ground state PEPS. Thus it may also be helpful for preparing excited state in quantum computers, which we leave to future study.

\par\noindent\emph{\textbf{Note added}} ---  When finalizing this work, we
became aware of a related work using AD methods for diagram summations in PEPS~\cite{Ponsioen2023b}.

\par\noindent\emph{\textbf{Acknowledgments}} ---
We acknowledge discussions with Ying-Jer Kao, Tsuyoshi Okubo, Rico Pohle, and Jens Eisert. Authors also acknowledge the Advanced Study Group~(ASG) \textsl{Tensor Network Approaches to Many-Body Systems} organized by the Center for Theoretical Physics of Complex Systems~(PCS) of the Institute for Basic Science~(IBS) in Daejeon, Korea. Part of the calculation in this work was conducted in the Supercomputer Center of ISSP, the University of Tokyo. W.-L.T. is supported by the Center of Innovations for Sustainable Quantum AI~(JST Grant Number JPMJPF2221). H.-Y.L. and W.-L.T. were supported by the National Research Foundation of Korea(NRF) grant funded by the Korea government(MSIT) (No. 2020R1I1A3074769 and RS-2023-00220471). N.K. is supported by JSPS KAKENHI Grants No. JP19H01809 and No. JP23H01092. J.-Y.C. was supported by Open Research Fund Program of the State Key Laboratory of Low-Dimensional Quantum Physics (project No.~KF202207), Fundamental Research Funds for the Central Universities, Sun Yat-sen University (project No.~23qnpy60), a startup fund from Sun Yat-sen University, the Innovation Program for Quantum Science and Technology 2021ZD0302100, and National Natural Science Foundation of China (NSFC) (grant No.~12304186).
Part of the calculations reported were performed on resources provided by the Guangdong Provincial Key Laboratory of Magnetoelectric Physics and Devices, No.~2022B1212010008.
N.S. is supported by the European Union’s Horizon 2020 program through the ERC-CoG SEQUAM (Grant No.~863476), and was funded in part by the Austrian Science Fund FWF (Grants No.~P36305 and F7117).

\bibliography{draft}

\end{document}